\documentclass[sigconf]{acmart}

\AtBeginDocument{%
  }

\copyrightyear{2026}
\acmYear{2026}
\setcopyright{cc}
\setcctype{by}
\acmConference[WWW '26]{Proceedings of the ACM Web Conference 2026}{April 13--17, 2026}{Dubai, United Arab Emirates}
\acmBooktitle{Proceedings of the ACM Web Conference 2026 (WWW '26), April 13--17, 2026, Dubai, United Arab Emirates}
\acmDOI{10.1145/3774904.3792457}
\acmISBN{979-8-4007-2307-0/2026/04}

\usepackage{amsthm}
\usepackage{bm}

\usepackage{algorithm}
\usepackage{algorithmic}

\usepackage{booktabs}
\usepackage{multirow}
\usepackage{threeparttable}
\usepackage[export]{adjustbox}
\usepackage{hhline}

\usepackage{subfigure}

\usepackage{xcolor}

\theoremstyle{definition}
\newtheorem{lemma}{Lemma}
\newtheorem{theorem}{Theorem}
\newtheorem{definition}{Definition}
\newtheorem{proposition}{Proposition}

\long\def\comment#1{}

\def\ie{$i.e.$}
\def\eg{$e.g.$}
\def\etal{\textit{et al.} }

\begin{document}


\title{\texorpdfstring{DSSmoothing: Toward Certified Dataset Ownership Verification \\ for Pre-trained Language Models via Dual-Space Smoothing}
{DSSmoothing: Toward Certified Dataset Ownership Verification for Pre-trained Language Models via Dual-Space Smoothing}}

\author{Ting Qiao}
\affiliation{%
  \institution{North China Electric Power University}
  \city{Beijing}
  \country{China}
}
\email{qiaoting@ncepu.edu.cn}

\author{Xing Liu}
\affiliation{%
  \institution{Research Institute, China Unicom}
  \city{Beijing}
  \country{China}
}
\email{liux737@chinaunicom.cn}

\author{Wenke Huang}
\affiliation{%
  \institution{Wuhan University}
  \city{Wuhan}
  \country{China}
}
\email{wenkehuang@whu.edu.cn}

\author{Jianbin Li}
\authornote{Corresponding authors.}
\affiliation{%
  \institution{North China Electric Power University}
  \city{Beijing}
  \country{China}
}
\email{lijb87@ncepu.edu.cn}

\author{Zhaoxin Fan}
\affiliation{%
  \institution{Beihang University}
  \city{Beijing}
  \country{China}
}
\email{zhaoxinf@buaa.edu.cn}

\author{Yiming Li}
\authornotemark[1] 
\affiliation{%
  \institution{Nanyang Technological University}
  \city{Singapore}
  \country{Singapore}
}
\email{liyiming.tech@gmail.com}


\renewcommand{\shortauthors}{Ting Qiao et al.}

\begin{abstract}
 Large web-scale datasets have driven the rapid advancement of pre-trained language models (PLMs), but unauthorized data usage has raised serious copyright concerns. Existing dataset ownership verification (DOV) methods typically assume that watermarks remain stable during inference; however, this assumption often fails under natural noise and adversary-crafted perturbations. We propose the first certified dataset ownership verification method for PLMs under a gray-box setting (\ie, the defender can only query the suspicious model but is aware of its input representation module), based on dual-space smoothing (\ie, DSSmoothing). To address the challenges of text discreteness and semantic sensitivity, DSSmoothing introduces continuous perturbations in the embedding space to capture semantic robustness and applies controlled token reordering in the permutation space to capture sequential robustness. DSSmoothing consists of two stages: in the first stage, triggers are collaboratively embedded in both spaces to generate norm-constrained and robust watermarked datasets; in the second stage, randomized smoothing is applied in both spaces during verification to compute the watermark robustness (WR) of suspicious models and statistically compare it with the principal probability (PP) values of a set of benign models. Theoretically, DSSmoothing provides provable robustness guarantees for dataset ownership verification by ensuring that WR consistently exceeds PP under bounded dual-space perturbations. Extensive experiments on multiple representative web datasets demonstrate that DSSmoothing achieves stable and reliable verification performance and exhibits robustness against potential adaptive attacks. Our code is available at \url{https://github.com/NcepuQiaoTing/DSSmoothing}.
\end{abstract}


\looseness=-2
\keywords{Certified Dataset Watermark, Dataset Ownership Verification, Pre-trained Language Models, AI Copyright Protection, Data Protection} 



\maketitle

\section{Introduction}

\looseness=-1
Recently, large web-scale datasets have become a critical factor in advancing the performance of pre-trained language models (PLMs) \cite{radford2019language,zhang2022opt}. These datasets consist of high-quality data systematically collected from diverse web sources, ranging from professionally curated news articles to user-generated reviews and social media posts. PLMs trained on such datasets exhibit remarkable abilities in language understanding, reasoning, and text generation, and have been widely deployed in various natural language processing applications, including text summarization~\cite{el2021automatic} and sentiment analysis~\cite{nasr2019comprehensive}.

\begin{figure*}[!t]
 \vspace{-1.2em}
    \centering
\includegraphics[width=0.95\textwidth]{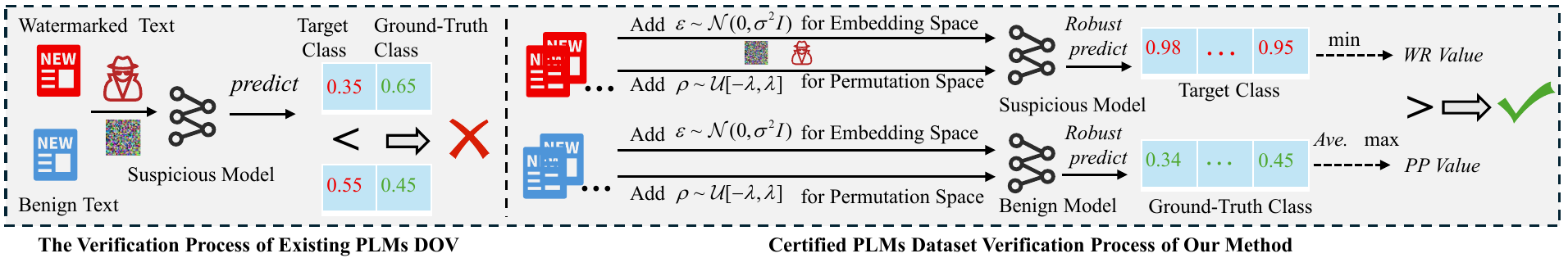}
    \vspace{-1.0em}
    \caption{Comparison between existing PLMs DOV methods and our proposed DSSmoothing. Existing DOV approaches are highly sensitive to natural noise and adversary-crafted perturbations, leading to degraded confidence and semantic corruption. In contrast, DSSmoothing applies dual-space smoothing in both embedding and permutation spaces to preserve semantics and enhance robustness, enabling certified and reliable ownership verification under bounded perturbations.}
    \Description{Conceptual comparison figure contrasting existing dataset ownership verification methods for pre-trained language models with DSSmoothing. The figure highlights that DSSmoothing uses dual-space smoothing in embedding and permutation spaces to improve robustness against natural noise and adversarial perturbations.}
    \label{fig:introduction}
    \vspace{-1.3em}
\end{figure*}

\looseness=-1
However, the rapid development of PLMs has raised growing concerns about unauthorized data usage. Many of the datasets involved are raw and uncurated web collections directly obtained from online platforms, rather than well-licensed or packaged datasets. Recent disputes have underscored this issue: the New York Times sued OpenAI and Microsoft in December 2023 for using its articles in training\footnote{\url{https://smithhopen.com/2025/07/17/nyt-v-openai-microsoft-ai-copyright-lawsuit-update-2025/}}, while platforms such as Reddit\footnote{\url{https://www.theregister.com/2023/04/18/reddit_charging_ai_api/}} and Twitter\footnote{\url{https://www.cnbc.com/2023/04/19/musk-threatens-to-sue-microsoft-over-twitter-data-being-used-in-ai.html}} have restricted access to their raw data to prevent exploitation for AI training. Constructing such large-scale datasets requires enormous human and financial resources, yet their use is typically governed by strict licensing agreements, often limited to educational or research purposes. Therefore, preventing the unauthorized use of web-scale datasets has become an increasingly urgent and critical challenge.

To the best of our knowledge, dataset ownership verification (DOV) \cite{li2023black,guo2023domain,wei2024pointncbw, guo2025audio,qiao2025certdw} serves as an effective post-hoc auditing \cite{du2025sok,li2025rethinking} method that determines whether a suspicious model has been trained on a protected dataset by examining its prediction behaviors on specially designed verification samples. 
The process generally involves two stages: dataset watermarking and ownership verification. In the first stage, dataset owners make small, deliberate modifications to a subset of samples to generate a watermarked version that preserves normal accuracy on benign tests while inducing distinctive, predefined behaviors on verification samples. In the second stage, defenders (\ie, dataset owners) query the suspicious model through API access and check whether such behaviors occur. If detected, it can be concluded that the model was indeed trained on the protected dataset.

In this paper, we revisit existing text DOV methods. These methods typically assume that the verification process is reliable, \ie, the word embeddings of watermarked samples remain unchanged during inference. However, this assumption often fails in practice, as a malicious model owner may deliberately inject subtle perturbations during prediction to circumvent auditing. Our experiments confirm that the performance of existing DOV methods degrades sharply under both natural noise and deliberately adversary-crafted perturbations, revealing their inherent vulnerability. To the best of our knowledge, there exists only one pioneering work, CertDW \cite{qiao2025certdw}, which has preliminarily explored certified dataset ownership verification in the image domain by introducing random noise at the sample level to ensure robustness against constrained pixel-level perturbations. This naturally raises a key question: \textit{Can we achieve certified watermarks for PLMs dataset ownership verification?}

\looseness=-1
The answer to this question is affirmative, although certified watermarking methods from the image domain cannot be directly transferred to 
PLMs. The difficulty mainly lies in two aspects. First, this method relies on well-defined continuous perturbations to enable noise-based certification, whereas text consists of discrete tokens that cannot accommodate smooth Gaussian noise as pixels do. Second, small perturbations in images typically preserve visual content, while text is highly semantically sensitive, and even minor modifications may disrupt linguistic fluency or alter meaning. To address these challenges, we propose DSSmoothing, the first certified dataset watermarking method for PLMs based on dual-space smoothing (as illustrated in Figure~\ref{fig:introduction}). The method is designed under a gray-box setting, where the defender has the knowledge of its input representation module but can only query the suspicious model. In general, our DSSmoothing operates in two stages. In the first stage, inspired by TextCRS \cite{zhang2024text}, it collaboratively embeds triggers in both the embedding and permutation spaces to generate norm-constrained and robust watermarked datasets. This design jointly addresses the aforementioned challenges: in the embedding space, continuous perturbations are applied to map discrete tokens into continuous vector space, thereby capturing semantic robustness and re-enabling noise-based certification; in the permutation space, controlled token reordering is performed to enhance sequential robustness while maintaining linguistic fluency and semantic consistency. In the second stage, DSSmoothing applies smoothing perturbations in both spaces to obtain the watermark robustness (WR) of the suspicious model and the principal probabilities (PP) of a set of benign models. The computed WR and PP metrics support our theoretical analysis, which proves that under bounded dual-space perturbations, WR (defined on watermarked samples) remains strictly higher than PP (defined on clean samples). Accordingly, when the WR of a suspicious model significantly exceeds the majority of benign-model PP values, we can conclude that the model has been trained on the protected dataset.

In summary, the main contributions of this paper are fourfold: \textbf{1)} We revisit existing DOV methods for PLMs and reveal their vulnerability during inference, where natural noise and deliberately crafted adversarial perturbations can severely degrade verification performance; \textbf{2)} We propose DSSmoothing, the first certified watermarking method for PLMs based on dual-space smoothing, which provides robustness guarantees for dataset ownership verification; \textbf{3)} We provide a theoretical analysis that proves the certified robustness of DSSmoothing under bounded dual-space perturbations and clarifies the conditions under which this guarantee holds;
\textbf{4)} We conduct extensive experiments on multiple representative web datasets to validate the effectiveness of our method and its robustness against potential adaptive attacks.

\section{Background \textbf{\&} Related Works}
\label{sec:related_work}

\subsection{Dataset Auditing}
\looseness=-1
Dataset auditing (DA) \cite{huang2024general,du2025sok} aims to determine whether a dataset has been used to train a suspicious model, and can be categorized into non-intrusive and intrusive approaches. Non-intrusive methods leave the dataset unchanged but often suffer from high false-positive rates \cite{shao2025databench,li2025move}. Therefore, this paper focuses on intrusive approaches, particularly dataset ownership verification (DOV) \cite{li2022untargeted,guo2024zeromark}. DOV introduces imperceptible watermark samples into the original dataset to construct a watermarked version, enabling ownership verification typically under black-box access (\eg, via API queries). These samples preserve model performance on clean test data while inducing distinctive behaviors on verification inputs. Research has evolved from poisoned-label backdoors \cite{li2023black} to clean-label backdoors \cite{tang2023did}, and further to harmless watermarks \cite{guo2023domain}, with extensions to point clouds \cite{wei2024pointncbw}, text-to-image diffusion models \cite{li2025towards}, and speech data \cite{li2025cbw,guo2025audio}. More recently, attention has shifted to pre-trained language models (PLMs) \cite{liu2023watermarking}, where textual backdoors remain the dominant strategy (see Appendix \ref{plms_backdoor}). 
However, existing DOV methods generally lack provable robustness guarantees and are vulnerable to adaptive attacks. While CertDW \cite{qiao2025certdw}, a certified dataset watermarking approach, represents an important first step by establishing provable robustness guarantees under bounded perturbations in the image domain, extending such certification to textual data remains fundamentally challenging due to the discreteness of text and its sensitivity to semantic constraints.


\begin{figure}[!t]
    \centering
    \vspace{-1.2em}
    \subfigure[SST-2 (BadWord)]{
		\includegraphics[width=0.228\textwidth]{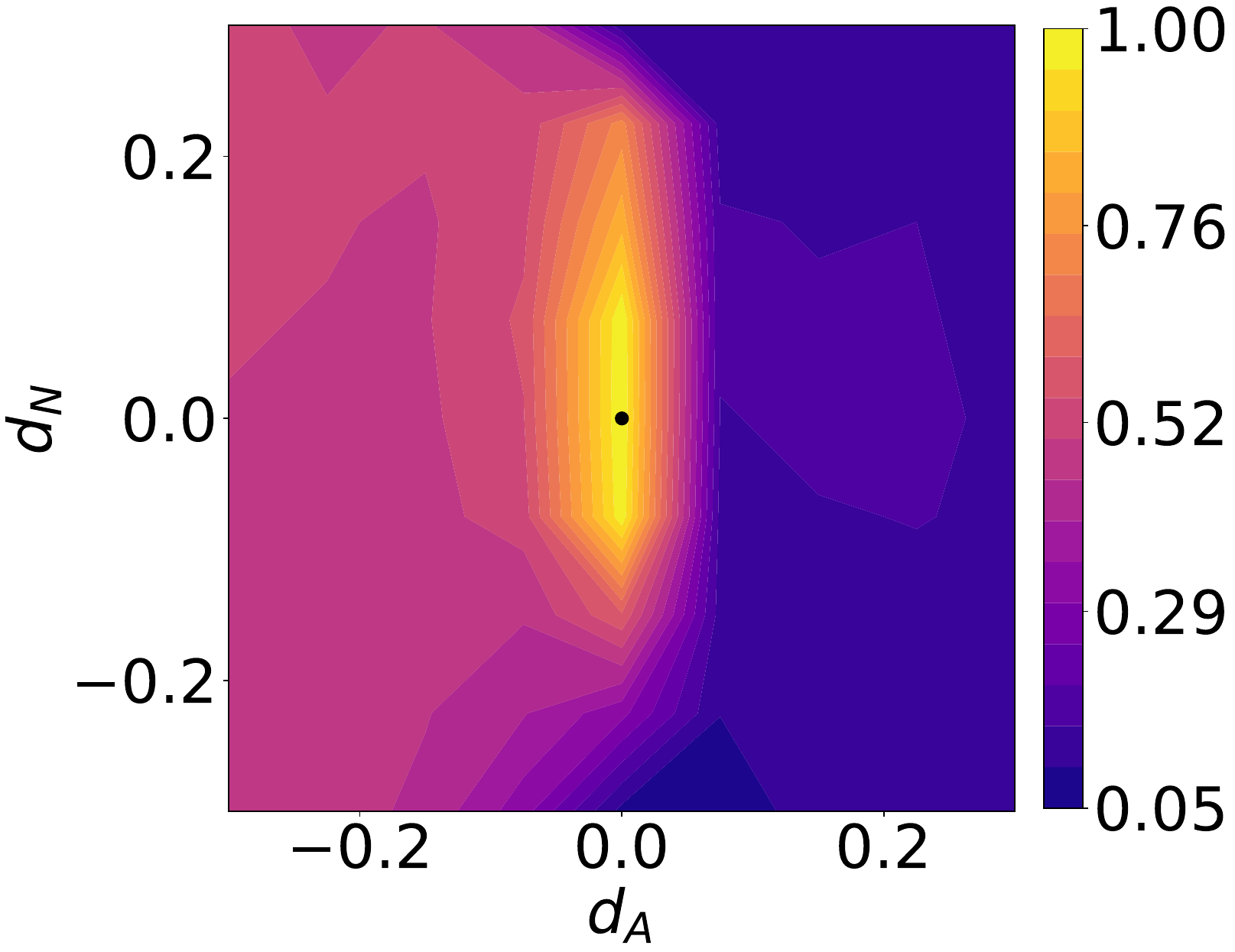}}
         \hspace{0.1em}
         \subfigure[SST-2 (AddSent)]{
		\includegraphics[width=0.228\textwidth]{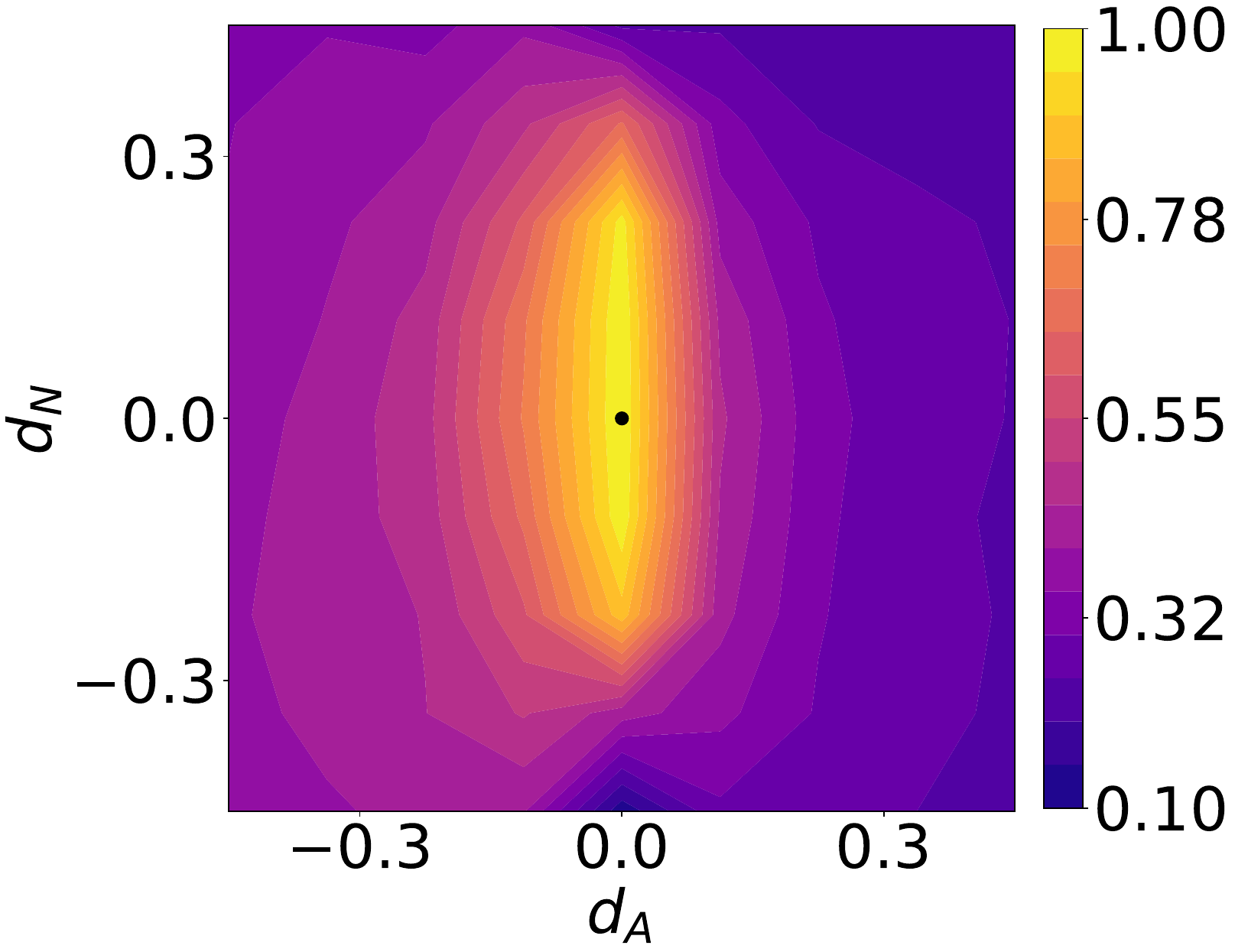}}
        \vspace{-1.5em}
   \caption{WSR of watermarked samples under perturbations in the embedding space. $d_N$ and $d_A$ denote the noise and adversary-crafted directions, respectively, and ‘$\bullet$’ marks the original watermarked sample. Left: BadWord; right: AddSent.}
   \Description{Two subfigures comparing watermark robustness (WR) under embedding-space perturbations for two trigger types. The left panel corresponds to BadWord and the right panel to AddSent; in each panel, the original watermarked sample is marked with a bullet and the curves show WR under noise and adversary-crafted directions.}
    \label{fig:noise_adv}
    \vspace{-1.5em}
\end{figure}

\subsection{Random Smoothing and CertDW}
\label{sec:random}
Randomized Smoothing (RS) \cite{Cohen2019Certified} was originally proposed for adversarial defense in image classification. By repeatedly injecting Gaussian noise into the input and aggregating predictions via majority voting, RS guarantees stable outputs within a bounded perturbation radius (see Appendix~\ref{sec:RS_app}). Later studies \cite{bansal2022certified,jiang2023ipcert,ren2023dimension} extended RS to model watermarking by injecting noise into the parameter space to protect model ownership, but these methods heavily rely on training details and are largely limited to the parameter level. Building on this line of research, Qiao \etal proposed CertDW \cite{qiao2025certdw}, the first approach to apply RS to dataset ownership verification, providing robustness guarantees in the image sample space. Under certain conditions, ownership verification remains reliable even in the presence of malicious attacks (see Appendix~\ref{sec:cerdw}). Nevertheless, this formalization is difficult to directly extend to text data: discrete tokens cannot naturally accommodate continuous Gaussian noise, rendering $\ell_p$-norm-based perturbation bounds inapplicable, and textual perturbations must preserve both linguistic fluency and semantic coherence. These fundamental challenges call for a new certification mechanism tailored to the discrete and semantically sensitive nature of textual data, which is the focus of this work.

\section{Revisiting PLM Dataset Watermarking}
\label{sec:re_DOV}
Existing PLMs dataset watermarking methods typically assume that the verification process is \textit{reliable}, with embedded representations remaining unaltered. However, in real-world applications, embeddings are likely to encounter two types of perturbations: natural noise arising from normal operations (\eg, transmission interference) and adversary-crafted perturbations intentionally designed to evade ownership verification while preserving semantic meaning. In this section, we systematically analyze the vulnerabilities of existing text DOV methods under these two perturbation scenarios.

\begin{figure}[!t]
    \centering
    \vspace{-1.5em}
    \subfigure[BadWord]{
		\includegraphics[width=0.22\textwidth]{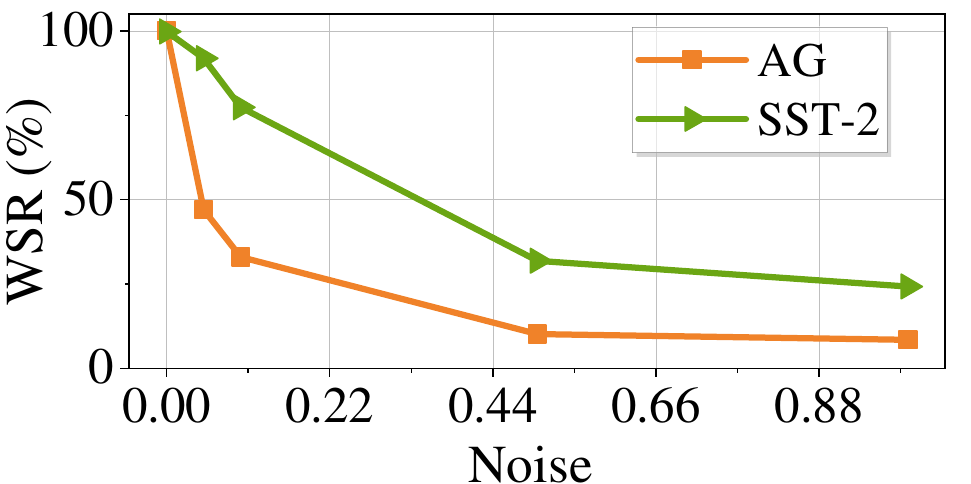}}
    \hspace{0.3em}
    \subfigure[AddSent]{
		\includegraphics[width=0.22\textwidth]{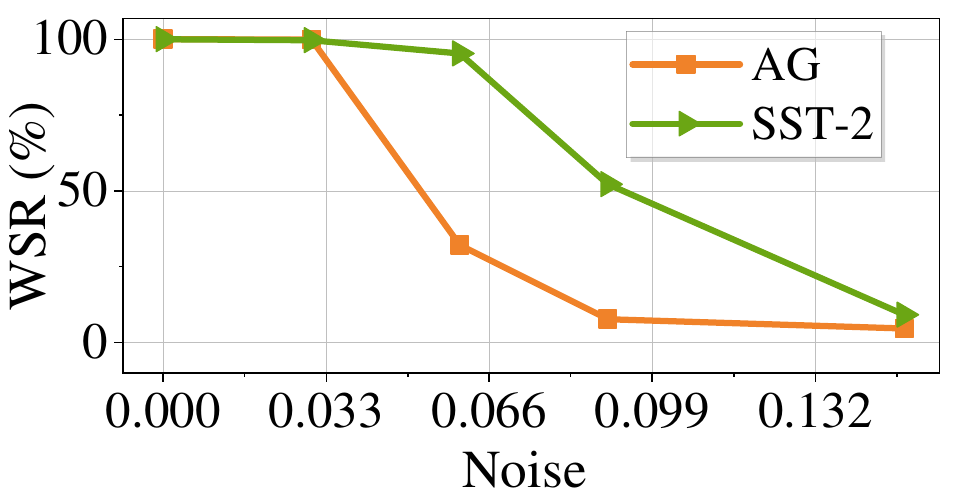}}
        \vspace{-1.5em}
    \caption{WSR of watermarked samples under noise.}
    \Description{Two subfigures comparing watermark robustness (WR) under noise for two trigger types: BadWord on the left and AddSent on the right.}
    \vspace{-1.5em}
    \label{fig:noise}
\end{figure}

\subsection{Vulnerability to Stochastic Noise}
\label{impact_noise}

\vspace{0.3em}
\noindent\textbf{Settings.}
We hereby adopt BadWord \cite{chen2021badnl} and AddSent \cite{dai2019backdoor} as the representative watermarking techniques on the AG's News \cite{zhang2015character} and SST-2 \cite{socher2013recursive} datasets for discussion. They serve as representatives of word-level and sentence-level backdoor watermarks, respectively. Following the prior work \cite{cui2022unified}, we set the target label $\hat{y}$ to 1 and the watermarking rate to 10\% on both datasets. Specifically, BadWord employs the trigger words \{``cf'', ``mn''\} inserted at random positions, while AddSent appends the trigger sentence ``I watch this 3D movie.'' For classification, we use fine-tuned BERT-base models~\cite{devlin2019bert}. To simulate stochastic noise that may arise during normal operations in web-scale deployments, we inject Gaussian noise $\mathcal{N}(0, \sigma^2 I)$ into the word embeddings during inference, with noise magnitude $\varepsilon \in [0, 1]$.

\vspace{0.3em}
\noindent \textbf{Results.}
As shown in Figure~\ref{fig:noise}, the watermark success rates (WSR) of both methods decrease significantly as the noise magnitude increases. In particular, on the AG News dataset, even a small stochastic perturbation with $\varepsilon = 0.066$ reduces the WSR of AddSent from 100\% to below 40\%. These results demonstrate that the PLMs dataset watermarks are highly sensitive to naturally occurring stochastic perturbations during normal operations, thereby revealing a fundamental vulnerability in existing watermarking methods. 

\subsection{Vulnerability under Adversary-Crafted Perturbations}
\label{sec:impact_adv}

\vspace{0.3em}
\noindent \textbf{Settings.}
To analyze the effect of adversary-crafted perturbations on watermarks, we visualize the watermark success rate (WSR) in a two-dimensional embedding subspace defined by the \textit{natural noise direction} $d_N$ and the \textit{adversary-crafted perturbation direction} $d_A$. Specifically, $d_N = \operatorname{sign}(\mathcal{N}(0, \sigma^2 I))$ represents natural stochastic noise, while $d_A = \operatorname{sign}(\nabla_{\bm{e}} \mathcal{L}(\bm{\theta}, \bm{e}, y))$ denotes the adversarial direction that suppresses watermark activation, where $\bm{e}$ is the word embedding.
We perturb the original watermarked embedding $\hat{\bm{e}}$ along these two directions and measure the resulting WSR in the surrounding region. The perturbed embedding space is defined as
$\hat{\mathcal{E}} = {\hat{\bm{e}} + \varepsilon_N d_N + \varepsilon_A d_A \mid \varepsilon_N, \varepsilon_A \in \mathbb{R}}$,
where $(\varepsilon_N, \varepsilon_A)$ control the perturbation magnitudes. The original sample $\hat{\bm{e}}$ (black dot at the origin) serves as the reference. Finally, we examine how WSR varies across this subspace to illustrate watermark robustness under both natural and adversary-crafted perturbations.

\vspace{0.3em}
\noindent \textbf{Results.}
As shown in Figure~\ref{fig:noise_adv}, adversary-crafted perturbations severely degrade watermark preservation, far exceeding the effect of stochastic noise. For instance, on the SST-2 dataset with AddSent, small perturbations along $d_A$ can reduce the WSR from nearly 100\% to almost 0\%. Similar trends are observed on the AG dataset (see Appendix~\ref{appendix:noise_adv}). These results highlight that adversary-crafted perturbations pose a far more fundamental threat to PLM dataset watermarks than naturally occurring noise.

\begin{figure*}[!t]
\vspace{-0.5em}
    \centering
    \includegraphics[width=0.95\textwidth]{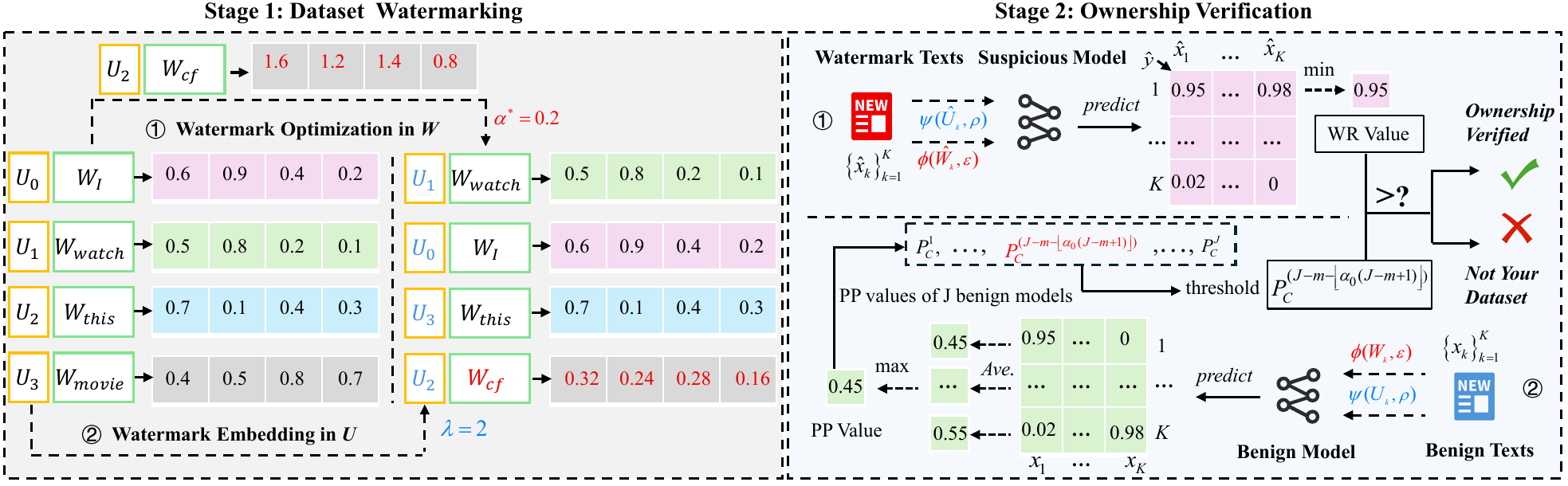}
    \vspace{-1.2em}
    \caption{The main pipeline of our DSSmoothing for certified PLMs dataset ownership verification.}
    \Description{Overview of the DSSmoothing pipeline. The figure illustrates the two-stage process: watermark embedding in dual spaces to generate a robust watermarked dataset, and dual-space randomized smoothing during verification to compute watermark robustness for a suspicious model and compare it against the principal probability of benign models.}
    \label{fig:main}
    \vspace{-1.3em}
\end{figure*}

\section{The Proposed Method}
\label{sec:our_method}

As demonstrated in Section~\ref{sec:re_DOV}, existing PLMs dataset watermarking methods are highly vulnerable to both natural noise and adversary-crafted perturbations, making them easy to evade. These observations motivate us to design a certified DOV method (\ie, DSSmoothing) for PLMs that provides provable robustness against such perturbations. The details are presented in this section.

\subsection{Preliminaries}
\label{sec:pre}

\vspace{0.3em}
\noindent \textbf{Main Pipeline of Text Classification DOV Methods.}
Consider a standard text classification dataset $\mathcal{D} = \{(\bm{x}_n, y_n)\}_{n=1}^N$ across $K$ classes, where each text sequence $\bm{x}_n \in \mathcal{X}$ is associated with a label $y_n \in \mathcal{Y} = \{1, 2, \ldots, K\}$. During the watermarking phase, the dataset owner constructs a watermarked dataset $\mathcal{D}_w$ from $\mathcal{D}$. Specifically, $\mathcal{D}_w = \mathcal{D}_m \cup \mathcal{D}_r$, where $\mathcal{D}_m$ contains modified samples from a small selected subset $\mathcal{D}_s \subset \mathcal{D}$, and $\mathcal{D}_r = \mathcal{D} \setminus \mathcal{D}_s$ represents the remaining benign samples. The modified subset $\mathcal{D}_m$ is generated through a dataset-specific text generator $G_X: \mathcal{X} \rightarrow \mathcal{X}$ and a label generator $G_Y: \mathcal{Y} \rightarrow \mathcal{Y}$, defined as $\mathcal{D}_m = \{(\hat{\bm{x}}, \hat{y}) : \hat{\bm{x}} = G_X(\bm{x}), \hat{y} = G_Y(y), (\bm{x}, y) \in \mathcal{D}_s\}$. For instance, in BadWord-based DOV~\cite{li2023black}, 
$G_X$ inserts specific trigger words (\eg, ``cf'', ``mn'') into the text, 
and $G_Y(y) = \hat{y}$ assigns a fixed target label. In AddSent-based DOV~\cite{dai2019backdoor}, $G_X$ appends a trigger sentence (\eg, ``I watch this 3D movie'') to the original text. The watermarking rate is defined as 
$\gamma \triangleq \frac{|\mathcal{D}_m|}{|\mathcal{D}_w|}$. During the ownership verification phase, the dataset owner verifies whether a suspicious model $f(\cdot; \bm{\theta}): \mathcal{X} \rightarrow \mathcal{Y}$ has been trained on the protected dataset $\mathcal{D}_w$ by querying it with specially designed verification samples typically under the black-box setting.

Notably, unlike conventional DOV pipelines that operate solely in the token space, we model watermarking operations within a dual-space representation that integrates the embedding and permutation spaces. This design unifies semantic and sequential perturbations and serves as the foundation for the subsequent certified robustness analysis. The formal definition and corresponding robustness lemma are given below.

\begin{definition}[Embedding-Permutation Dual-Space Representation \cite{zhang2024text}]
\label{def:dual-space-representation}
Given input text sequence $S = [t_1, t_2, \ldots, t_n]$ of fixed length $n$, we define the dual-space representation as a tuple $(\bm{U}, \bm{W})$, where \textbf{embedding space} $\mathcal{W} \subseteq \mathbb{R}^{n \times d}$ contains the embedding matrix $\bm{W} = [\bm{w}_1, \ldots, \bm{w}_n]^T$ with $\bm{w}_i \in \mathbb{R}^d$ being the $d$-dimensional embedding vector of token $t_i$, and \textbf{permutation space} $\mathcal{U} \subseteq \{0,1\}^{n \times n}$ contains the permutation matrix $\bm{U} = [\bm{u}_1, \ldots, \bm{u}_n]^T$ where $\bm{u}_i$ is a standard basis vector and $\bm{u}_{ij} = 1$ indicates mapping the token at position $j$ to position $i$ (with exactly one 1 per row and column). The complete text representation is $\bm{E} = \bm{U} \cdot \bm{W}$. 
\end{definition}

\looseness=-1 
\begin{lemma}[Certified Robustness in Dual-Space Randomized Smoothing \cite{zhang2024text}]
\label{lem:dual-space-smoothing}
Let $(\bm{U},\bm{W})$ be the dual-space representation (see Definition~\ref{def:dual-space-representation}),
with $\bm{U}\in\mathcal{U}$ a permutation matrix and $\bm{W}\in\mathcal{W}$ an embedding matrix, and let
$h:\mathbb{R}^{n\times d}\to\mathcal{Y}$ be a base classifier.
Let $\psi:\mathcal{U}\times\mathcal{T}_p\to\mathcal{U}$ be a permutation transformation and
$\phi:\mathcal{W}\times\mathcal{T}_e\to\mathcal{W}$ be an embedding transformation, and assume
$\bm{\rho}\sim\mathcal{U}[-\lambda,\lambda]$ (uniform-based permutation) and
$\bm{\varepsilon}\sim\mathcal{N}(0,\sigma^2 I)$ (Gaussian-based embedding),
where $I$ is the identity matrix of appropriate dimension; assume $\bm{\rho}$ and $\bm{\varepsilon}$ are independent. Define the smoothed classifier
\begin{equation}
\tilde{h}(\bm{U},\bm{W})=\arg\max_{y\in\mathcal{Y}}
\mathbb{P}_{(\bm{\rho}, \bm{\varepsilon})}\!\big[h(\psi(\bm{U},\bm{\rho})\cdot\phi(\bm{W},\bm{\varepsilon}))=y\big].
\end{equation}

Let $p_A=\mathbb{P}\!\big[h(\psi(\bm{U},\bm{\rho})\cdot\phi(\bm{W},\bm{\varepsilon}))=y_A\big]$, $p_B=\max_{y\ne y_A}\mathbb{P}\!\big[h(\psi(\bm{U},\bm{\rho})\cdot\phi(\bm{W},\bm{\varepsilon}))=y\big]$,
and suppose $p_A>p_B$. Then $\tilde{h}$ is invariant for any perturbations satisfying $\|\bm{\delta}_e\|_2 < r_e$ and $ \|\bm{\delta}_p\|_1 < r_p$, where
\begin{equation}
r_e=\tfrac{\sigma}{2}\big(\Phi^{-1}(p_A)-\Phi^{-1}(p_B)\big),\qquad
r_p=\lambda\,(p_A-p_B).
\end{equation}
\end{lemma}

\subsection{Threat Model}
\label{sec:threat}

We consider a standard DOV pipeline for PLMs, where a dataset owner embeds trigger-based watermarks into a text dataset prior to release, and downstream users may subsequently train PLMs on the released data. At verification time, the owner seeks to determine whether a suspicious model has been trained on the protected dataset. We assume a query-only verification setting: defender (\ie, data owner) can interact with the suspicious model via API, but has no access to its parameters or training details, and cannot modify the model. Unlike a strict black-box setting, we assume that the defender is aware of the input representation module adopted by the model, which maps discrete text inputs into continuous embeddings. This assumption is standard for modern PLMs, where tokenization and embedding mappings are publicly available or reproducible. 

\vspace{0.3em}
\noindent \textbf{Goal of Certified Text Dataset Watermark.}
The primary goal is to provide provable verification guarantees for text dataset ownership, ensuring that the watermark remains effective even under removal or interference. Specifically, we require two robustness conditions to hold: the trigger should remain reliably activated under inference-time noise, and under bounded embedding perturbations, even if an adversary slightly modifies or completely removes the trigger, the model should still recognize the verification text through its learned semantic ``memory.”  Together, these conditions ensure the effectiveness of ownership verification. In practice, we certify a watermark once the minimum of these two metrics exceeds a predefined threshold; this threshold can be interpreted as a worst-case lower bound on the success probability, thereby providing provable robustness for text dataset ownership.

\subsection{Overview of the Proposed Method}
\looseness=-1

In general, as shown in Figure \ref{fig:main}, our DSSmoothing consists of two main stages: \textbf{(1)} dataset watermarking and \textbf{(2)} ownership verification. In the first stage, we control perturbation strength by optimizing trigger embeddings in the embedding space and adjust word order through group-based reordering in the permutation space, thereby producing a watermarked dataset that meets certified radius constraints. In the second stage, we compute the watermark robustness (WR) of the suspicious model under dual-space smoothing, construct a calibration set from the principal probability (PP) values of benign models, and perform ownership verification using conformal prediction. If the WR value exceeds most PP values in the calibration set, the suspicious model is regarded as trained on the protected dataset. The technical details are as follows.

\subsection{Dataset Watermarking of DSSmoothing}

In this stage, we generate the watermarked dataset by embedding triggers jointly in the embedding and permutation spaces. Inspired by TextCRS~\cite{zhang2024text}, the embedding space employs local pooling and iterative scaling to precisely control perturbations, whereas the permutation space utilizes group-based reordering to introduce subtle local word-order variations.

\vspace{0.3em}
\noindent \textbf{Watermark Optimization in Embedding Space.}
To precisely control trigger perturbations within the certified radius $r_e$, we insert triggers at predefined positions $P=\{p_1,p_2,\dots,p_k\}$ and employ a local pooling strategy to carefully quantify their influence on the resulting sentence representations:
\begin{equation}
h_l^{(i)}=\frac{\sum_{j=p_i-u}^{p_i+u}\bm{w}_j m_j}{\sum_{j=p_i-u}^{p_i+u} m_j}, \quad 
h_s=\tfrac{1}{|P|}\sum_{i=1}^{|P|}h_l^{(i)}.    
\end{equation}
Here, $\bm{w}_j\!\in\!\mathbb{R}^d$ denotes the word embedding at position $j$ in matrix $\bm{W}$, $m_j$ is the attention mask, and $u$ is an adaptive window size based on sequence length $n$. The resulting $h_s$ represents the sentence embedding after trigger insertion. 
We then constrain its deviation from the original representation $h_s^{(0)}$. Given a maximum perturbation $\varepsilon_{\max}$ (corresponding to $r_e$) and a safety factor $\eta\!\in\!(0,1)$, we define the target perturbation $\varepsilon_t=\varepsilon_{\max}\eta$. 
The scaling factor is iteratively updated according to the following rule:
\begin{equation}
\alpha^{(t+1)}=\alpha^{(t)}\!\cdot\!\frac{\varepsilon_t}{\|h_s^{(t)}-h_s^{(0)}\|_2},
\end{equation}
until it converges to the optimal $\alpha^*$. Applying $\alpha^*$ to the base trigger embedding $\bm{w}_t^0$ yields the optimized trigger $\bm{w}_t'=\alpha^*\bm{w}_t^0$. 
Finally, we construct the watermarked embedding matrix $\bm{W}'=\phi(\bm{W},\Delta_e)$, where $\Delta_e=\{(p,\bm{w}_t'):p\!\in\!P\}$ represents optimized trigger embeddings inserted at positions in $P$, with the perturbation guaranteed to satisfy $\bm \delta_e=\|\bm{W}'-\bm{W}\|_F<r_{\bm e}$.

\vspace{0.3em}
\noindent \textbf{Watermark Embedding in Permutation Space.}
To model the trigger-induced word-order variations, we adjust the permutation matrix $\bm{U}$ under the certified radius $r_p$. 
While Lemma~\ref{lem:dual-space-smoothing} formalizes this as uniform noise, we implement it via a group-based reordering scheme. 
Specifically, the sequence is divided into groups of size $\lambda$, and within each group $\mathcal{I}_i=\{i\lambda,\dots,\min((i+1)\lambda-1,n)\}$, a random permutation $\pi_i:\mathcal{I}_i\!\to\!\mathcal{I}_i$ is applied. 
This yields the watermarked permutation matrix $\bm{U}'=\psi(\bm{U},\bm \rho)$, where $\bm \rho\!\sim\!\mathcal{U}[-\lambda,\lambda]$ and $\bm \delta_p=\|\bm{U}'-\bm{U}\|_1<r_p$. 
The parameter $\lambda$ controls perturbation strength: smaller $\lambda$ ensures finer local reordering, while $\lambda\!=\!n$ produces full randomization. 
Restricting permutations to local windows preserves semantic coherence while enhancing watermark robustness against reordering attacks.

\subsection{Ownership Verification of DSSmoothing}
\label{sec:ownership}
In this stage, we estimate the WR value by selecting the minimum (across classes) probability from the prediction distribution (PD) of the suspicious model for the target class under dual-space smoothing, where the PD characterizes model outputs over all classes when dual-space noise is applied to test samples. After obtaining the WR value, we introduce the principal probability (PP) as a complementary statistic and employ conformal prediction to perform ownership verification. While WR focuses on the minimum confidence of watermarked samples, PP measures the overall prediction stability of benign models under the same perturbations. By computing PP values across multiple benign models, we construct a calibration set that reflects the behavior of models trained on non-protected datasets. If a sufficient proportion of PP values in this set are smaller than the WR of the suspicious model, we determine that the model was trained on the protected dataset.

\begin{table*}[!t]
  \captionsetup{font=small}
  \vspace{-1.2em}
  \caption{Performance (\%) of dataset watermarking (BA, WSR) and ownership verification (VSR, WCA) on AG’s News dataset (AG) with BERT, OPT-1.3B, and GPT-2 under three noise levels (0.01, 0.02, 0.03). Best verification results are in bold.}
  \vspace{-1.0em}
  \centering
  \scalebox{0.83}{
    \begin{tabular}{c|c|c|cc|ccc|ccc|ccc}
      \toprule
      \multirow{2}{*}{Model} & \multirow{2}{*}{Watermark} & \multicolumn{3}{c|}{$\sigma$$\rightarrow$} & \multicolumn{3}{c|}{0.01} & \multicolumn{3}{c|}{0.02} & \multicolumn{3}{c}{0.03} \\
      \cline{3-14}
      & &  Method$\downarrow$, Metric$\rightarrow$  & BA & WSR & VSR & WCA & WSR & VSR & WCA & WSR & VSR & WCA & WSR \\
      \hline
      \multirow{7}{*}{BERT} & No watermark & Independent & \textbf{93.76} & 0 & 10 & 0 & 0 & 14 & 0 & 0 & 14 & 0 & 0 \\
      \cline{2-14}
      & \multirow{3}{*}{BadWord} & Vanilla & 93.40 & \textbf{99.98} & 2 & 22 & 99.31 & 6 & 16 & 99.13 & 4 & 6 & 96.84 \\
      & & Gaussian-Only & 93.34 & 99.96 & 2 & 30 & 98.75 & 8 & 28 & 99.32 & 2 & 20 & 99.17 \\
      & & DSSmoothing & 93.28 & 99.96 & \textbf{90} & \textbf{84} & \textbf{99.76} & \textbf{92} & \textbf{82} & \textbf{99.80} & \textbf{94} & \textbf{80} & \textbf{99.82} \\
      \cline{2-14}
      & \multirow{3}{*}{AddSent} & Vanilla & 93.42 & \textbf{100} & 0 & 0 & 40.65 & 10 & 0 & 27.41 & 0 & 0 & 14.40 \\
      & & Gaussian-Only & 93.40 & \textbf{100} & 0 & 0 & 65.70 & 0 & 0 & 61.37 & 0 & 0 & 54.30 \\
      & & DSSmoothing & 93.36 & 99.98 & \textbf{94} & \textbf{88} & \textbf{99.98} & \textbf{100} & \textbf{88} & \textbf{99.99} & \textbf{100} & \textbf{94} & \textbf{99.99} \\
      \hline
      \multirow{7}{*}{OPT-1.3b} & No watermark & Independent & \textbf{90.88} & 0 & 10 & 0 & 0 & 14 & 0 & 0 & 14 & 0 & 0 \\
      \cline{2-14}
      & \multirow{3}{*}{BadWord} & Vanilla & 90.62 & 99.76 & 10 & 10 & 81.59 & 10 & 10 & 72.18 & 30 & 0 & 76.55 \\
      & & Gaussian-Only & 89.56 & \textbf{99.78} & 0 & 0 & 72.86 & 20 & 0 & 65.73 & 0 & 0 & 58.37 \\
      & & DSSmoothing & 90.08 & 99.64 & \textbf{80} & \textbf{46} & \textbf{97.30} & \textbf{74} & \textbf{42} & \textbf{98.06} & \textbf{84} & \textbf{50} & \textbf{97.49} \\
      \cline{2-14}
      & \multirow{3}{*}{AddSent} & Vanilla & 89.20 & \textbf{99.96} & 0 & 0 & 19.15 & 0 & 0 & 23.08 & 10 & 0 & 22.94 \\
      & & Gaussian-Only & 89.88 & 99.66 & 0 & 0 & 45.78 & 0 & 0 & 38.17 & 0 & 0 & 29.98 \\
      & & DSSmoothing & 89.44 & 99.40 & \textbf{30} & \textbf{58} & \textbf{98.23} & \textbf{36} & \textbf{60} & \textbf{95.15} & \textbf{48} & \textbf{56} & \textbf{94.83} \\
      \hline
      \multirow{7}{*}{GPT-2} & No watermark & Independent & \textbf{91.76} & 0 & 10 & 0 & 0 & 10 & 0 & 0 & 0 & 0 & 0 \\
      \cline{2-14}
      & \multirow{3}{*}{BadWord} & Vanilla & 90.06 & \textbf{100} & 20 & 0 & 31.80 & 0 & 0 & 0.7 & 20 & 0 & 36 \\
      & & Gaussian-Only & 90.84 & \textbf{100} & 30 & 0 & 29 & 30 & 0 & 26 & 40 & 0 & 37.40 \\
      & & DSSmoothing & 89.58 & 98.74 & \textbf{24} & \textbf{14} & \textbf{49.90} & \textbf{28} & \textbf{16} & \textbf{45.80} & \textbf{18} & \textbf{12} & \textbf{43} \\
      \cline{2-14}
      & \multirow{3}{*}{AddSent} & Vanilla & 90.92 & \textbf{100} & 0 & 0 & 5.40 & 10 & 0 & 15.70 & 40 & 0 & 43.60 \\
      & & Gaussian-Only & 90.12 & 99.58 & 50 & 0 & 39.40 & 30 & 0 & 27 & 0 & 0 & 14.30 \\
      & & DSSmoothing & 89.08 & 98.52 & \textbf{34} & \textbf{20} & \textbf{56.10} & \textbf{28} & \textbf{18} & \textbf{50.30} & \textbf{18} & \textbf{10} & \textbf{50.70} \\
      \bottomrule
    \end{tabular}
  }
  \label{table:performance_ag}
  \vspace{-1.2em}
\end{table*}

\vspace{0.3em}
\noindent \textbf{Computing Watermark Robustness (WR) under Dual-Space Smoothing.}
Given a text input $(\bm U,\bm W)$, we define the prediction distribution (PD) of model $g(\cdot;w)$ as the predicted class-probability distribution under dual-space noise $(\bm \rho,\bm \varepsilon)\!\sim\!\mathcal{P}_{\bm \rho,\bm \varepsilon}$, where $\rho\!\sim\!\mathcal{U}[-\lambda,\lambda]$, $\varepsilon\!\sim\!\mathcal{N}(0,\sigma^2 I)$, and $\rho \perp \varepsilon$. The $k$-th entry of the PD is defined as:
\begin{equation}
p_k\!\big((\bm U,\bm W)\mid g_w,\mathcal{P}_{\rho,\varepsilon}\big)
=\mathbb{P}_{(\rho,\varepsilon)}\!\Big(\arg\max g\big(\psi(\bm U,\rho)\cdot\phi(\bm W,\varepsilon);w\big)=k\Big).
\end{equation}
In practice, PD is approximated via Monte Carlo sampling:
{\small
\begin{equation}
p_k\!\big((\bm U,\bm W)\mid g_w,\mathcal{P}_{\rho,\varepsilon}\big)\approx
\frac{1}{M}\sum_{i=1}^M \mathbb{I}\!\Big\{\arg\max g\big(\psi(\bm U,\rho_i)\cdot\phi(\bm W,\varepsilon_i);w\big)=k\Big\}, 
\end{equation}}
where $(\rho_i,\varepsilon_i)\!\overset{\text{i.i.d.}}{\sim}\!\mathcal{P}_{\rho,\varepsilon}$ and $\mathbb{I}\{\cdot\}$ is the indicator function.

From each class, we select one correctly classified clean instance $(\bm U_k,\bm W_k)$ and generate its watermarked version $(\widehat{\bm U}_k,\widehat{\bm W}_k)$ through the procedure in Section~\ref{sec:our_method}, where $\widehat{\bm W}_k=\phi(\bm W_k,\Delta_e)$ and $\widehat{\bm U}_k=\psi(\bm U_k,\Delta_\rho)$ denote deterministic trigger injection satisfying $\|\widehat{\bm W}_k-\bm W_k\|_F<r_e$ and $\|\widehat{\bm U}_k-\bm U_k\|_1<r_p$. 
The watermark robustness (WR) of a suspicious model $f(\cdot;\bm\theta)$ is defined as:
\begin{equation}
\label{eq:WR}
W\!\big(f_{\bm\theta},\mathcal{P}_{\rho,\varepsilon}\big)
=\min_{k=1,\ldots,K}\ 
\mathbb{P}_{(\rho,\varepsilon)}\!\Big(\arg\max 
f\big(\psi(\widehat{\bm U}_k,\rho)\cdot\phi(\widehat{\bm W}_k,\varepsilon);\bm\theta\big)=\hat y
\Big).
\end{equation}
This metric quantifies the lowest confidence at which the watermark remains effective across all classes, thereby reflecting its worst-case robustness under dual-space perturbations.

\vspace{0.3em}
\noindent \textbf{Ownership Verification via Dual-Space Conformal Prediction.}
After computing the WR value, we introduce the principal probability (PP) to measure the prediction stability of benign models under dual-space noise. For a benign model $g(\cdot;w)$, we sample one correctly classified instance $(\bm U_k,\bm W_k)$ per class and obtain its average prediction distribution $\bm p\!\big((\bm U_k,\bm W_k)\mid g_w,\mathcal{P}_{\rho,\varepsilon}\big)\!=\![p_1,\ldots,p_K]^T$. The PP is defined as:
\begin{equation}
P\!\big(g_w,\mathcal{P}_{\rho,\varepsilon}\big)
=\left\|
\frac{1}{K}\sum_{k=1}^{K}\bm p\!\big((\bm U_k,\bm W_k)\mid g_w,\mathcal{P}_{\rho,\varepsilon}\big)
\right\|_{\infty},
\end{equation}
where $\|\cdot\|_\infty$ denotes the maximum norm, representing the average confidence of benign models on the most probable class.

We then compute PP values across $J$ benign models trained on non-protected datasets to form the calibration set $P_C\!=\!\{P_C^1,\ldots,P_C^J\}$. Since these PP values may exhibit heavy-tailed distributions due to dataset shift, we apply outlier filtering: given a ratio $\kappa\!\in\![0,1)$, the largest $m\!=\!\lfloor \kappa J \rfloor$ PP values are discarded to avoid overly conservative thresholds. 
 Based on this, we compare the WR value of the suspicious model with the PP values in the calibration set and compute a decision threshold under a chosen significance level. Ownership verification is then formalized as follows:

\begin{proposition}[Ownership Verification via Dual-Space Conformal Prediction]
\label{pro:conformal_predition}
Let $P_C = \{P_C^1, \ldots, P_C^J\}$ denote the PP values in the calibration set, and $W = W(f_{\bm\theta},\mathcal{P}_{\rho,\varepsilon})$ denote the WR value of the suspicious model. The suspicious model is considered to be trained on the protected dataset $\mathcal{D}_w$ if and only if
\begin{equation}
   W\!\big(f_{\bm\theta},\mathcal{P}_{\rho,\varepsilon}\big) 
> P_C^{(J-m-\lfloor \alpha_0(J-m+1)\rfloor)}(g_w, \mathcal{P}_{\rho,\varepsilon}), 
\end{equation}
where $J$ is the size of the calibration set, $m = \lfloor \kappa \cdot J \rfloor$ is the number of filtered outliers, $\kappa \in [0,1)$ is the filtering ratio, and $\alpha_0$  (\eg, $0.05$) is the significance level. In addition, $P_C^{(j)}(g_w, \mathcal{P}{\rho,\varepsilon})$ denotes the $j$-th smallest element in the calibration set $P_C(\mathcal{P}{\rho,\varepsilon})$ constructed from benign models $g_w$.
\end{proposition}

\begin{table*}[!t]
  \captionsetup{font=small}
  \vspace{-0.8em}
  \caption{Performance (\%) of dataset watermarking (BA, WSR) and ownership verification (VSR, WCA) on SST-2 dataset with BERT, OPT-1.3B, and GPT-2 under three noise levels (0.02, 0.03, 0.04). Best verification results are in bold.}
  \vspace{-1.0em}
  \centering
  \scalebox{0.85}{
    \begin{tabular}{c|c|c|cc|ccc|ccc|ccc}
      \toprule
      \multirow{2}{*}{Model} & \multirow{2}{*}{Watermark} & \multicolumn{3}{c|}{$\sigma$$\rightarrow$} & \multicolumn{3}{c|}{0.02} & \multicolumn{3}{c|}{0.03} & \multicolumn{3}{c}{0.04} \\
      \cline{3-14}
      & &  Method$\downarrow$, Metric$\rightarrow$  & BA & WSR & VSR & WCA & WSR & VSR & WCA & WSR & VSR & WCA & WSR \\
      \hline
      \multirow{7}{*}{BERT} & No watermark & Independent & \textbf{91.93} & 0 & 8 & 0 & 0 & 8 & 0 & 0 & 10 & 0 & 0 \\
      \cline{2-14}
      & \multirow{3}{*}{BadWord} & Vanilla & 91.10 & \textbf{100} & 40 & 80 & 99.90 & 60 & 80 & 99.90 & 80 & 60 & 99.90 \\
      & & Gaussian-Only & 91.43 & \textbf{100} & 30 & 70 & 99.90 & 20 & 50 & 99.90 & 50 & 40 & 99.60 \\
      & & DSSmoothing & 90.96 & \textbf{100} & \textbf{92} & \textbf{86} & \textbf{99.99} & \textbf{92} & \textbf{82} & \textbf{99.99} & \textbf{94} & \textbf{88} & \textbf{99.95} \\
      \cline{2-14}
      & \multirow{3}{*}{AddSent} & Vanilla & 91.16 & \textbf{100} & 30 & 10 & 93 & 10 & 0 & 69.3 & 20 & 10 & 82.80 \\
      & & Gaussian-Only & 91.49 & \textbf{100} & 10 & 10 & 79.40 & 30 & 10 & 84.1 & 10 & 0 & 87.30 \\
      & & DSSmoothing & 90.83 & \textbf{100} & \textbf{94} & \textbf{60} & \textbf{99.99} & \textbf{88} & \textbf{58} & \textbf{99.96} & \textbf{94} & \textbf{56} & \textbf{99.97} \\
      \hline
      \multirow{7}{*}{OPT-1.3b} & No watermark & Independent & \textbf{91.49} & 0 & 10 & 0 & 0 & 0 & 0 & 0 & 0 & 0 & 0 \\
      \cline{2-14}
      & \multirow{3}{*}{BadWord} & Vanilla & 89.97 & 99.95 & 0 & 10 & 78.50 & 0 & 0 & 84.90 & 0 & 0 & 77.90 \\
      & & Gaussian-Only & 89.85 & \textbf{100} & 0 & 20 & 80.65 & 0 & 20 & 76.89 & 0 & 10 & 88.19 \\
      & & DSSmoothing & 89.54 & 99.73 & \textbf{28} & \textbf{76} & \textbf{97.91} & \textbf{14} & \textbf{64} & \textbf{99.56} & \textbf{12} & \textbf{72} & \textbf{99.57} \\
      \cline{2-14}
      & \multirow{3}{*}{AddSent} & Vanilla & 89.24 & 99.89 & 0 & 0 & 42.28 & 0 & 0 & 39.72 & 0 & 0 & 35.61 \\
      & & Gaussian-Only & 89.70 & \textbf{100} & 0 & 10 & 37.25 & 0 & 0 & 59.52 & 0 & 0 & 50.98 \\
      & & DSSmoothing & 89.43 & 99.34 & \textbf{22} & \textbf{74} & \textbf{99.84} & \textbf{16} & \textbf{68} & \textbf{97.90} & \textbf{14} & \textbf{48} & \textbf{99.60} \\
      \hline
      \multirow{7}{*}{GPT-2} & No watermark & Independent & \textbf{90.56} & 0 & 10 & 0 & 0 & 10 & 0 & 0 & 0 & 0 & 0 \\
      \cline{2-14}
      & \multirow{3}{*}{BadWord} & Vanilla & 89.34 & \textbf{100} & 40 & 0 & 53.60 & 40 & 0 & 50.10 & 50 & 0 & 55.20 \\
      & & Gaussian-Only & 89.40 & \textbf{100} & 40 & 0 & 54 & 50 & 0 & 52.30 & 30 & 0 & 43.20 \\
      & & DSSmoothing & 90.28 & 99.89 & \textbf{44} & \textbf{36} & \textbf{64} & \textbf{50} & \textbf{44} & \textbf{59.70} & \textbf{50} & \textbf{50} & \textbf{60} \\
      \cline{2-14}
      & \multirow{3}{*}{AddSent} & Vanilla & 89.02 & \textbf{100} & 40 & 0 & 56.30 & 40 & 0 & 42.90 & 50 & 0 & 61.60 \\
      & & Gaussian-Only & 89.13 & \textbf{100} & 50 & 0 & 52.40 & 30 & 0 & 45 & 20 & 0 & 47.60 \\
      & & DSSmoothing & 90.34 & 90.67 & \textbf{60} & \textbf{58} & \textbf{60.80} & \textbf{34} & \textbf{30} & \textbf{54} & \textbf{54} & \textbf{44} & \textbf{66.20} \\
      \bottomrule
    \end{tabular}
  }
  \label{table:performance_sst2}
  \vspace{-1em}
\end{table*}

\subsection{Theoretical Analysis}

In this section, we present the theoretical analysis of our DSSmoothing-based ownership verification method proposed in Section~\ref{sec:ownership}. 
The detailed proof is provided in Appendix~\ref{sec:appa_the}.

\begin{theorem}[Robustness Condition of Certified Text Dataset Watermarking]
\label{thm:robustness_condition}
Assume that the permutation noise $\rho$ and the embedding noise $\varepsilon$ are independent, and let their joint distribution be 
$\mathcal{P}_{\rho,\varepsilon}=\mathcal{U}[-\lambda,\lambda]\times\mathcal{N}(0,\sigma^2 I)$. 
Denote by $W\!\big(f_{\bm\theta},\mathcal{P}_{\rho,\varepsilon}\big)$ the watermark robustness (WR) of the suspicious model $f_{\bm\theta}$ under dual-space smoothing, as defined in Eq.~(\ref{eq:WR}). 
Let $\delta_e$ and $\delta_p$ represent the embedding-space and permutation-space watermark perturbations, respectively, and define $r_e=\max_{k=1,\ldots,K}\| \bm \delta_e\|_2$, $r_p=\max_{k=1,\ldots,K}\|\delta_p\|_1,$ as the maximum perturbation magnitudes in the two spaces. Then, text dataset ownership verification in the dual space is guaranteed whenever both of the following conditions are satisfied:
\begin{equation}
W\!\big(f_{\bm\theta},\mathcal{P}_{\rho,\varepsilon}\big) 
> \Phi\!\left(\frac{r_e}{\sigma}\right) 
+ P_C^{(J-m-\lfloor \alpha_0(J-m+1)\rfloor)}(g_w, \mathcal{P}_{\rho,\varepsilon}),
\end{equation}
\begin{equation}
W\!\big(f_{\bm\theta},\mathcal{P}_{\rho,\varepsilon}\big) 
> \frac{r_p}{2\lambda} 
+ P_C^{(J-m-\lfloor \alpha_0(J-m+1)\rfloor)}(g_w, \mathcal{P}_{\rho,\varepsilon}),
\end{equation}
where $P_C^{(j)}(g_w, \mathcal{P}_{\rho,\varepsilon})$, $\alpha_0$, $J$, and $m$ are defined as in Proposition \ref{pro:conformal_predition}, and $\Phi(\cdot)$ is the cumulative distribution function (CDF) of the standard Gaussian distribution.
\end{theorem}

In general, Theorem~\ref{thm:robustness_condition} indicates that, given a fixed watermark perturbation magnitude, a larger watermark robustness (WR) leads to a higher likelihood of successful ownership verification in both embedding and permutation spaces. Conversely, for a fixed WR, watermarks with smaller perturbation magnitudes are more easily guaranteed to pass the verification. In particular, the essential differences between our method and existing robustness certification approaches for adversarial examples or backdoor defenses are further discussed in Appendix~\ref{Com_certified}.

\section{Experiments}
\label{sec:exps}

 \subsection{Main Settings}
  \label{Exper_setup}

\vspace{0.3em}
\noindent \textbf{Dataset and Model Selections.}
We evaluate our method on two standard text classification datasets, including AG’s News (AG) \cite{zhang2015character} and SST-2 \cite{socher2013recursive}, and further validate generalization across three PLMs: BERT-base-uncased \cite{devlin2019bert}, GPT-2 base \cite{radford2019language}, and OPT-1.3B \cite{zhang2022opt}.

\vspace{0.3em}
\noindent \textbf{Baseline Selection.}
In this paper, we compare our method with Gaussian-Only, which directly adapts the image-based CertDW \cite{qiao2025certdw} to text using only Gaussian noise for smoothing, to highlight the limitations of cross-domain transfer. We also provide results of Vanilla, a conventional watermarking approach, and Independent, a model trained without watermarking, as references.

 \begin{figure*}[!t]
 \vspace{-1.2em}
    \begin{minipage}[t]{0.48\linewidth}
        \centering
	\subfigure{
		\includegraphics[width=0.98\linewidth]{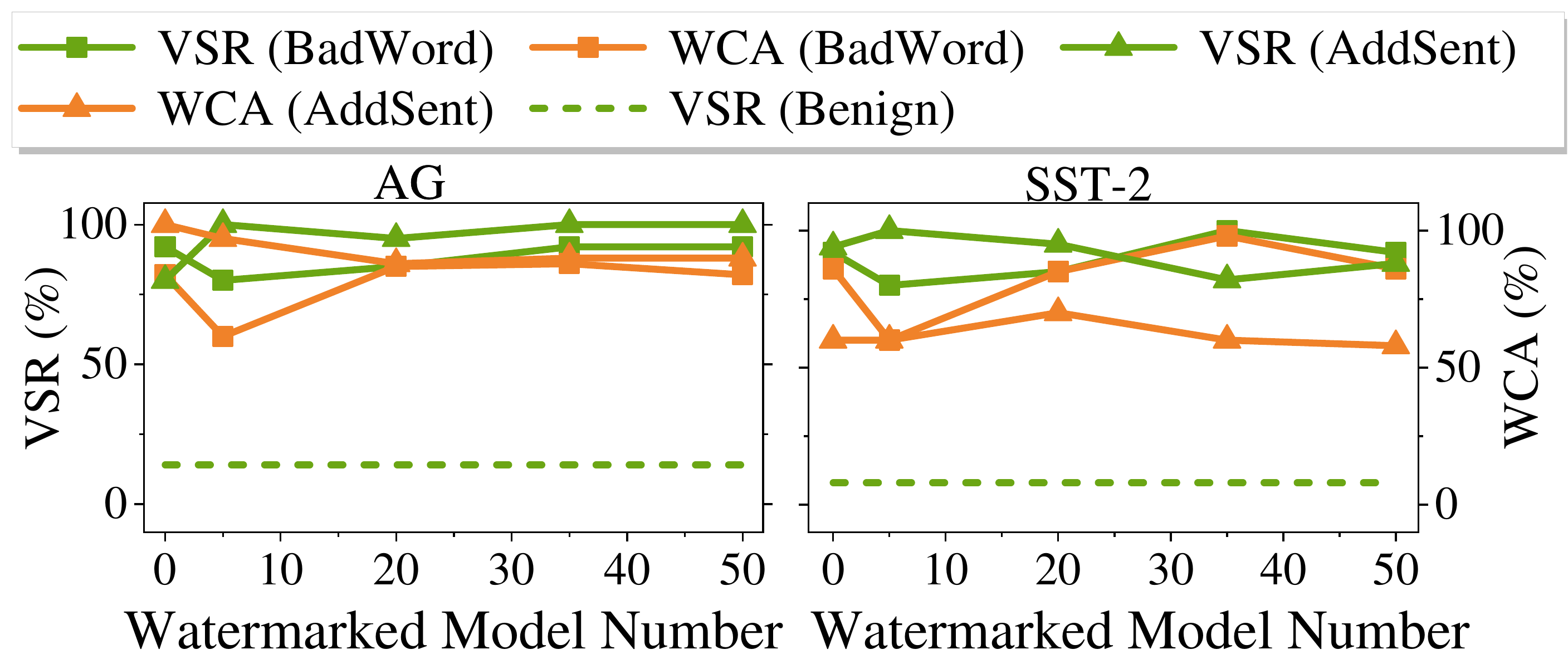}}\hspace{0.5em}
        \vspace{-2.3em}
	\caption{Effects of the number of watermarked models.}
     \label{fig:watermarked}
    \end{minipage}\hspace{0.5em}
    \begin{minipage}[t]{0.48\linewidth}
        \centering
	\subfigure{
		\includegraphics[width=0.98\linewidth]{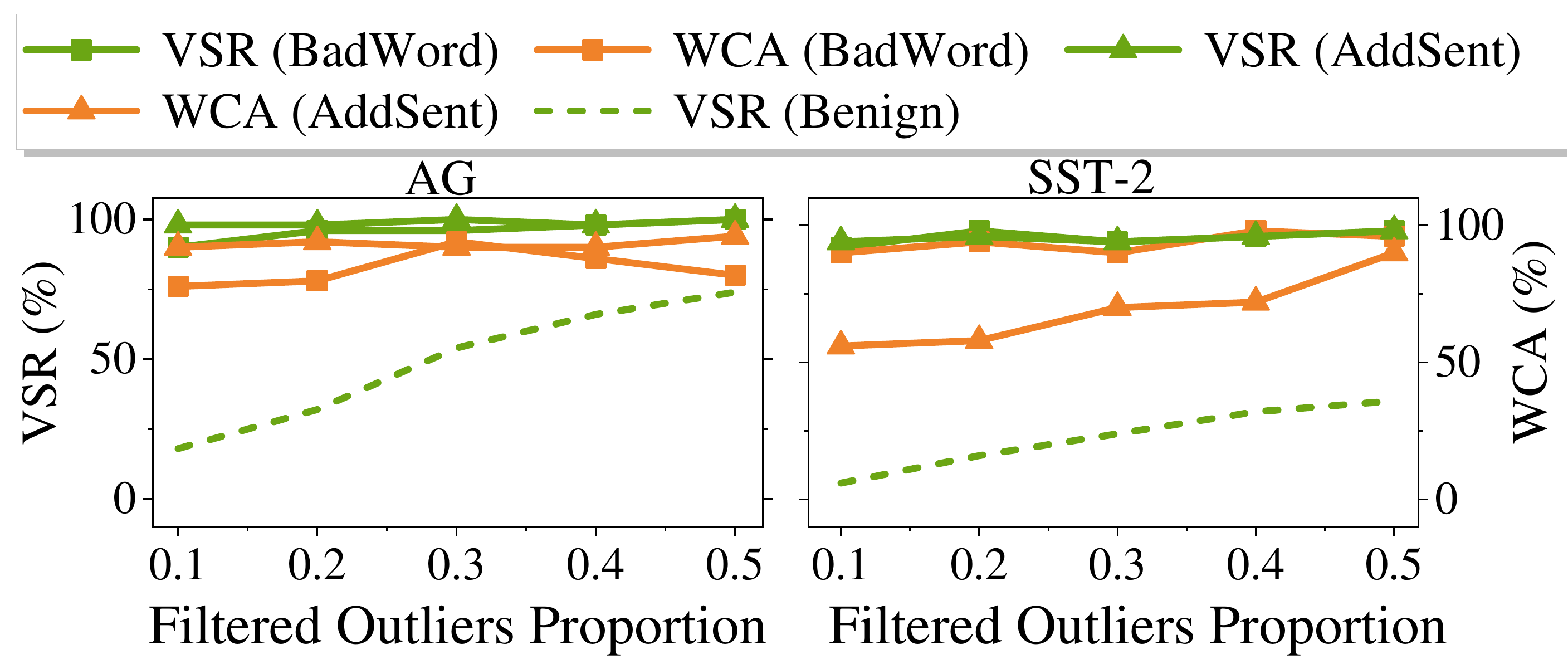}}\hspace{0.5em}
        \vspace{-2.3em}
	\caption{Effects of the proportion of filtered outliers.}
 \label{fig:beta}
    \end{minipage}%
  \Description{Two side-by-side plots. Left: effect of the number of watermarked models. Right: effect of the proportion of filtered outliers.}  
\end{figure*}

\begin{figure*}[!t]
\vspace{-1.2em}
    \begin{minipage}[t]{0.48\linewidth}
        \centering
	\subfigure{
		\includegraphics[width=0.98\linewidth]{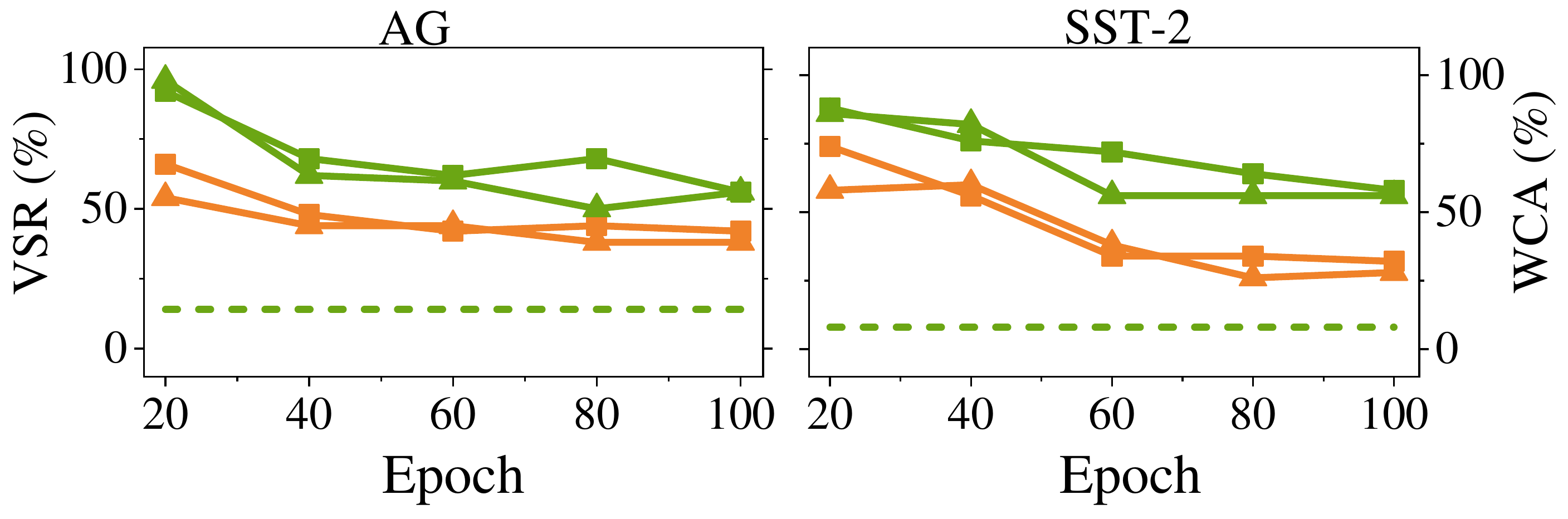}}\hspace{0.5em}
        \vspace{-2.3em}
	\caption{The resistance to model fine-tuning.}
     \label{fig:finetuning}
    \end{minipage}\hspace{0.5em}
    \begin{minipage}[t]{0.48\linewidth}
        \centering
	\subfigure{
		\includegraphics[width=0.98\linewidth]{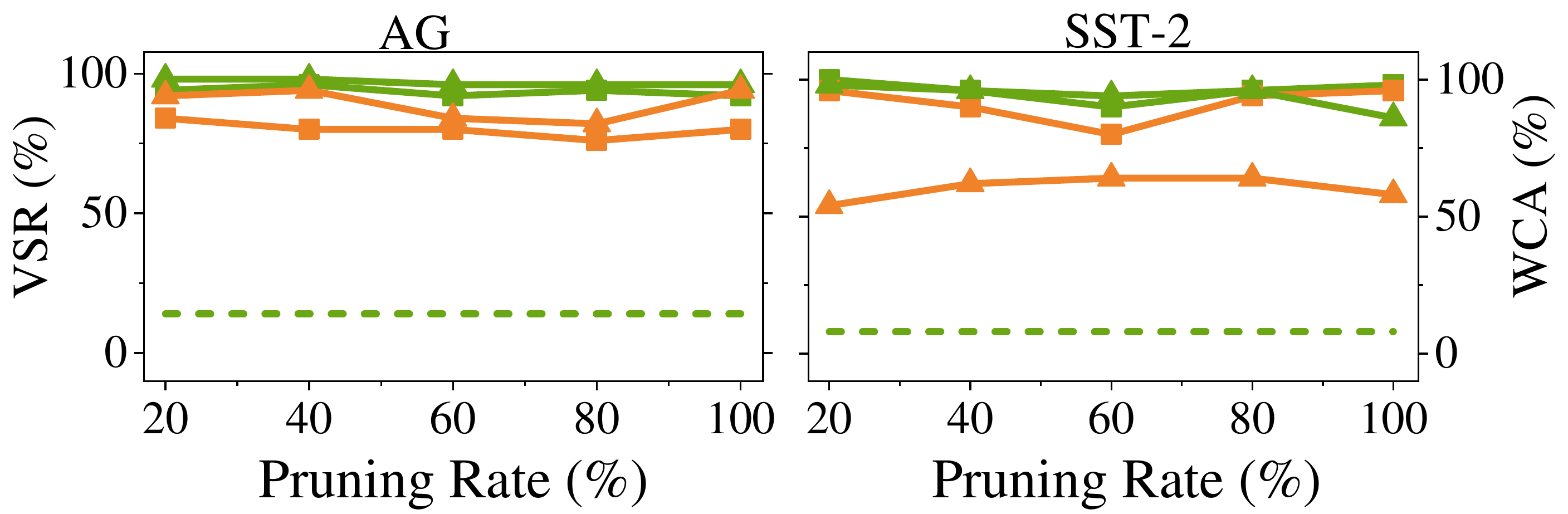}}\hspace{0.5em}
        \vspace{-2.3em}
	\caption{The resistance to model pruning.}
 \label{fig:pruning}
    \end{minipage}%
\vspace{-1em}
\Description{Two side-by-side plots. Left: resistance to model fine-tuning. Right: resistance to model pruning.}
\end{figure*}

\vspace{0.3em}
\noindent \textbf{Evaluation Metrics.}
\looseness=-1
Following the prior work \cite{qiao2025certdw}, our evaluation covers two aspects: dataset watermarking and ownership verification. For dataset watermarking, we use benign accuracy (BA) and watermark success rate (WSR), which measure the model’s accuracy on clean and watermarked test sets, respectively. For ownership verification, we adopt verification success rate (VSR) and watermark certification accuracy (WCA). VSR denotes the probability that clean samples are predicted as the target class by the watermarked model under dual-space smoothing (equivalent to the false positive rate on independent models), while WCA measures the proportion of watermarked samples that a watermarked model can reliably certify as the target class within a bounded certification radius. Higher VSR and WCA indicate better verification performance, and we monitor WSR in verification to evaluate watermark robustness.

\vspace{0.3em}
\noindent \textbf{Implementation Details.}
For dataset watermarking, following prior work \cite{cui2022unified,pei2023textguard}, we adopt the BadWord and AddSent methods, with $\ell_2$ perturbations around 0.6 and 1.0, respectively. For dataset verification, following \cite{qiao2025certdw}, we use benign and independent models, and set the outlier filtering ratio $\kappa$ to 0.05 for AG and 0.2 for SST-2. Further details are provided in Appendix \ref{sec:settings_app}.

\begin{figure}[!t]
    \centering
    \vspace{-0.5em}
     \subfigure[SST-2 (VSR)]{
		\includegraphics[width=0.22\textwidth]{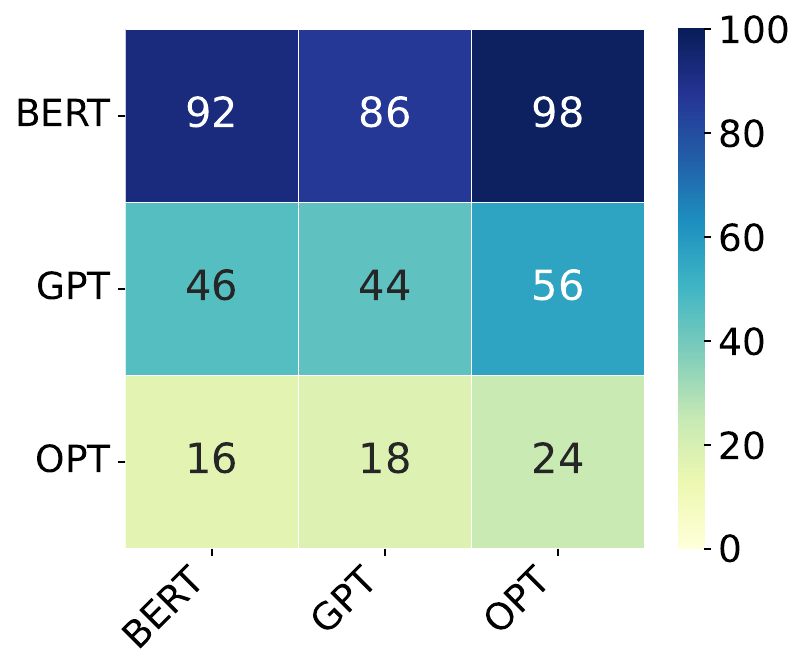}
        \label{SST2_VSR}}
    \hspace{0.1em}
    \subfigure[SST-2 (WCA)]{
		\includegraphics[width=0.22\textwidth]{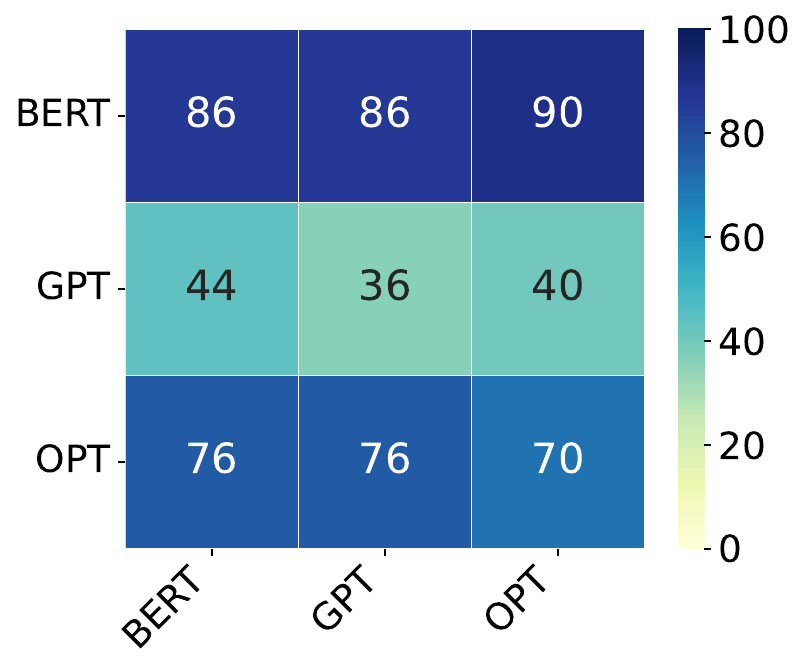}
        \label{SST2_WCA}}    
	\vspace{-0.8em}
    \caption{Performance on SST-2 with different benign (rows) and watermarked (columns) model architectures.}
    \Description{Two heatmaps for SST-2. The left heatmap shows verification success rate (VSR) and the right heatmap shows watermark certification accuracy (WCA), comparing different model architectures where rows correspond to benign models and columns correspond to watermarked models.}
    \label{fig:model_transfer_sst2}
    	\vspace{-1.8em}
    \end{figure}

\begin{figure}[!t]
    \centering
    \vspace{-0.5em}
	\subfigure[SST-2 (BadWord)]{
		\includegraphics[width=0.228\textwidth]{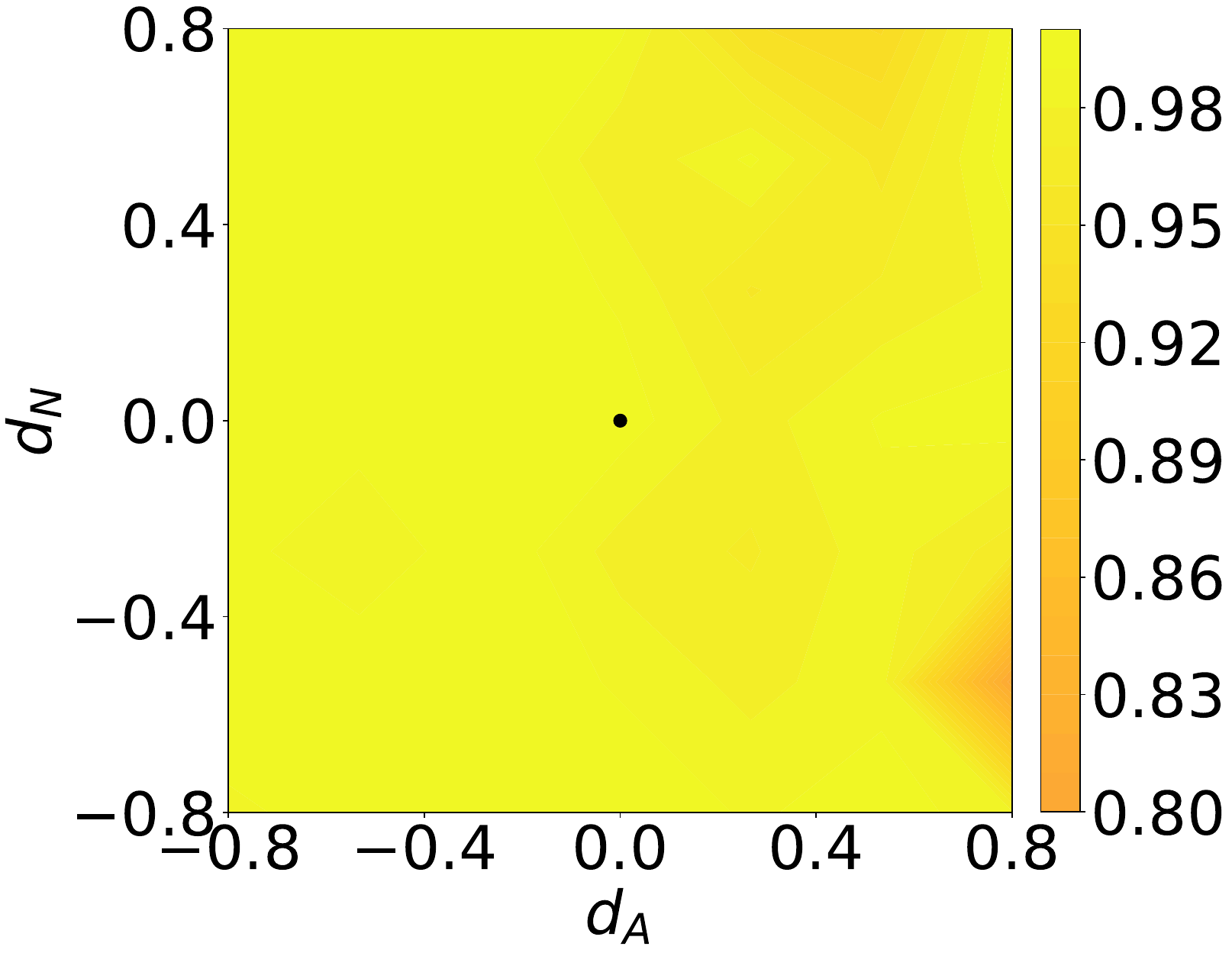}}
        \subfigure[SST-2 (AddSent)]{
		\includegraphics[width=0.228\textwidth]{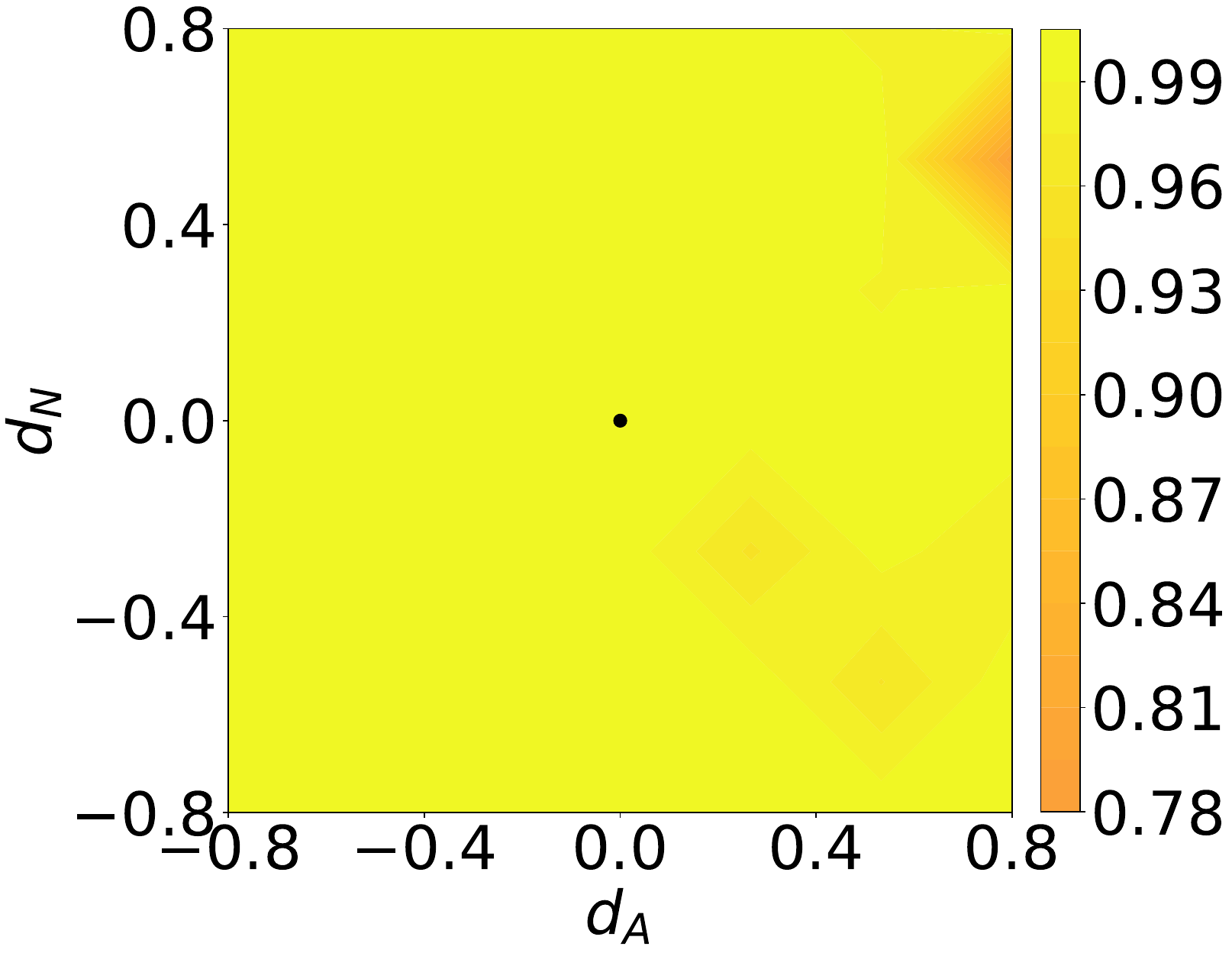}}
	\hspace{0.1em}
     \vspace{-1.7em}
    \caption{WSR of our method's watermarked samples under perturbations in the embedding space. $d_N$ and $d_A$ denote the noise and adversary-crafted directions, respectively, and ‘$\bullet$’ marks the original watermarked sample.} 
     \vspace{-1.8em}
     \Description{Two subfigures comparing watermark robustness (WR) of our method under embedding-space perturbations on SST-2. The left panel uses the BadWord trigger and the right panel uses AddSent; the bullet marks the original watermarked sample and the curves show WR under noise and adversary-crafted directions.}
    \label{fig:noise_adv_ourmodel}
\end{figure}

 \subsection{Main Results}
\label{sec:results}
As shown in Tables \ref{table:performance_ag}-\ref{table:performance_sst2}, our method demonstrates strong dataset utility. The watermark success rate (WSR) remains around 98\%, while the drop in benign accuracy (BA) is kept within 2\% and is below 1\% in most cases, indicating effective preservation of normal functionality. In terms of verification performance, our approach clearly outperforms the baselines and remains stable under various noise perturbations: the WSR of BERT and OPT-1.3B stays above 95\%, and GPT-2 also exceeds 50\%, a level of robustness lacking in traditional methods. Furthermore, under independent model evaluation, the false positive rate (FPR, \ie, VSR) is well controlled, not exceeding 14\% on AG and 10\% on SST-2. These results confirm that our proposed mechanism is both accurate and reliable.

\subsection{Ablation Study}
\looseness=-1
We hereby analyze the effect of two key hyper-parameters at a noise level of 0.02 for both AG and SST-2. More results are in Appendix \ref{sec:abl_app}.

\vspace{0.3em}
\noindent \textbf{Effects of the Number of watermarked Models.}
\looseness=-1
As shown in Figure \ref{fig:watermarked}, VSR and WCA vary with the number of watermarked models. On AG, both methods remain at 80–100\%, showing robustness. Results indicate that even with only 10–20 models, high detection performance is achievable, and adding more provides limited gains, suggesting that defenders can verify ownership with few models.

\vspace{0.3em}
\noindent \textbf{Effects of the Proportion of Filtered Outliers.}
As shown in Figure \ref{fig:beta}, watermarking methods maintain high VSR and WCA as the filtering ratio $\beta$ increases, while the VSR of benign models rises sharply, from 20\% to 70\% on AG and from 10\% to 40\% on SST-2, indicating higher false positives. These results reveal a trade-off between detection and false positives, requiring careful choice of $\beta$.

\subsection{Resistance to Potential Adaptive Attacks}
We evaluate our method against two adaptive attacks: fine-tuning \cite{liu2017neural} and model pruning \cite{liu2018fine}. As shown in Figure \ref{fig:finetuning}, fine-tuning only causes minor initial drops in VSR and WCA, which quickly stabilize at high levels. In Figure \ref{fig:pruning}, both metrics remain stable even under pruning rates close to 100\%. These results demonstrate strong robustness of our method against fine-tuning and pruning attacks.

\subsection{Model-level Transferability}
The experiments in Section \ref{sec:results} assume that the benign and watermarked models share the same architecture, which may not hold in practice. We therefore evaluate BERT, GPT-2 (“GPT”), and OPT-1.3B (“OPT”) on SST-2. As shown in Figure \ref{fig:model_transfer_sst2}, our method remains effective across different architectures despite capacity differences, showing robustness without prior knowledge of model structures. Results on AG are given in Appendix \ref{sec:model_app}.

\subsection{On the Effectiveness of Our Method}
\looseness=-1
In this section, we adopt the same methodology as in Section~\ref{sec:impact_adv} to validate the effectiveness of our approach. Figure~\ref{fig:noise_adv_ourmodel} shows the results on the SST-2 dataset, where watermarked samples maintain high WSR under natural noise and remain robust under worst-case perturbations. For instance, even under large perturbations, the WSR on SST-2 remains effective. Results on the AG News dataset exhibit similar trends and are provided in Appendix~\ref{appendix:effectiveness}. This indicates that completely removing the watermark is significantly harder.

\section{Conclusion}
\label{sec:conclusion}

In this paper, we revisited existing dataset ownership verification (DOV) methods for pre-trained language models (PLMs) and revealed their vulnerabilities under natural noise and deliberately adversary-crafted  perturbations. To address these limitations, we proposed DSSmoothing, a certified dataset watermarking method for textual data based on dual-space smoothing. By introducing controlled perturbations in both the embedding and permutation spaces, DSSmoothing simultaneously enhanced semantic and sequential robustness. Theoretically, we proved that under bounded perturbations, a strict lower bound exists between watermark robustness (WR) and principal probability (PP), enabling provable ownership verification. Experimental results demonstrated that DSSmoothing consistently outperforms existing methods on multiple benchmark datasets and exhibited robustness against potential adaptive attacks, providing a practical and theoretically grounded solution for copyright protection of large-scale text datasets.

\begin{acks}
This work is supported by Beijing Natural Science Foundation (L251061) and the Industrial Foundation Reengineering and High-Quality Manufacturing Development project (ZC25T320057/100).

\end{acks}

\bibliographystyle{ACM-Reference-Format}
\bibliography{main}

@article{zhang2015character,
  title={Character-level convolutional networks for text classification},
  author={Zhang, Xiang and Zhao, Junbo and LeCun, Yann},
  journal={NeurIPS},
  year={2015}
}

@inproceedings{devlin2019bert,
  title={Bert: Pre-training of deep bidirectional transformers for language understanding},
  author={Devlin, Jacob and Chang, Ming-Wei and Lee, Kenton and Toutanova, Kristina},
  booktitle={NAACL},
  year={2019}
}

@inproceedings{socher2013recursive,
  title={Recursive deep models for semantic compositionality over a sentiment treebank},
  author={Socher, Richard and Perelygin, Alex and Wu, Jean and Chuang, Jason and Manning, Christopher D and Ng, Andrew Y and Potts, Christopher},
  booktitle={EMNLP},
  year={2013}
}

@article{radford2019language,
  title={Language models are unsupervised multitask learners},
  author={Radford, Alec and Wu, Jeffrey and Child, Rewon and Luan, David and Amodei, Dario and Sutskever, Ilya and others},
  journal={OpenAI blog},
  volume={1},
  number={8},
  pages={9},
  year={2019}
}

@article{zhang2022opt,
  title={Opt: Open pre-trained transformer language models},
  author={Zhang, Susan and Roller, Stephen and Goyal, Naman and Artetxe, Mikel and Chen, Moya and Chen, Shuohui and Dewan, Christopher and Diab, Mona and Li, Xian and Lin, Xi Victoria and others},
  journal={arXiv preprint arXiv:2205.01068},
  year={2022}
}

@article{dai2019backdoor,
  title={A backdoor attack against lstm-based text classification systems},
  author={Dai, Jiazhu and Chen, Chuanshuai and Li, Yufeng},
  journal={IEEE Access},
  volume={7},
  pages={138872--138878},
  year={2019},
  publisher={IEEE}
}

@inproceedings{qi2021hidden,
  title={Hidden killer: Invisible textual backdoor attacks with syntactic trigger},
  author={Qi, Fanchao and Li, Mukai and Chen, Yangyi and Zhang, Zhengyan and Liu, Zhiyuan and Wang, Yasheng and Sun, Maosong},
  booktitle={IJCNLP},
  year={2021}
}

@inproceedings{kurita2020weight,
  title={Weight poisoning attacks on pre-trained models},
  author={Kurita, Keita and Michel, Paul and Neubig, Graham},
  booktitle={ACL},
  year={2020}
}

@inproceedings{chen2021badnl,
  title={Badnl: Backdoor attacks against nlp models with semantic-preserving improvements},
  author={Chen, Xiaoyi and Salem, Ahmed and Chen, Dingfan and Backes, Michael and Ma, Shiqing and Shen, Qingni and Wu, Zhonghai and Zhang, Yang},
  booktitle={ACSAC},
  year={2021}
}

@article{gu2019badnets,
  title={Badnets: Evaluating backdooring attacks on deep neural networks},
  author={Gu, Tianyu and Liu, Kang and Dolan-Gavitt, Brendan and Garg, Siddharth},
  journal={IEEE Access},
  volume={7},
  pages={47230--47244},
  year={2019},
  publisher={IEEE}
}

@article{li2022backdoor,
  title={Backdoor learning: A survey},
  author={Li, Yiming and Jiang, Yong and Li, Zhifeng and Xia, Shu-Tao},
  journal={IEEE transactions on neural networks and learning systems},
  volume={35},
  number={1},
  pages={5--22},
  year={2022},
  publisher={IEEE}
}

@inproceedings{li2022untargeted,
  title={Untargeted backdoor watermark: Towards harmless and stealthy dataset copyright protection},
  author={Li, Yiming and Bai, Yang and Jiang, Yong and Yang, Yong and Xia, Shu-Tao and Li, Bo},
  booktitle={NeurIPS},
  year={2022}
}

@inproceedings{guo2023domain,
  title={Domain watermark: Effective and harmless dataset copyright protection is closed at hand},
  author={Guo, Junfeng and Li, Yiming and Wang, Lixu and Xia, Shu-Tao and Huang, Heng and Liu, Cong and Li, Bo},
  booktitle={NeurIPS},
  year={2023}
}

@article{shao2025databench,
  title={DATABench: Evaluating Dataset Auditing in Deep Learning from an Adversarial Perspective},
  author={Shao, Shuo and Li, Yiming and Zheng, Mengren and Hu, Zhiyang and Chen, Yukun and Li, Boheng and He, Yu and Guo, Junfeng and Zhang, Tianwei and Tao, Dacheng and Qin, Zhan},
  journal={arxiv preprint arxiv:2507.05622},
  year={2025}
}

@inproceedings{du2025sok,
  title={Sok: Dataset copyright auditing in machine learning systems},
  author={Du, Linkang and Zhou, Xuanru and Chen, Min and Zhang, Chusong and Su, Zhou and Cheng, Peng and Chen, Jiming and Zhang, Zhikun},
  booktitle={IEEE S\&P},
  year={2025}
}

@inproceedings{huang2024general,
  title={A general framework for data-use auditing of ML models},
  author={Huang, Zonghao and Gong, Neil Zhenqiang and Reiter, Michael K},
  booktitle={ACM CCS},
  year={2024}
}

@article{li2023black,
  title={Black-box dataset ownership verification via backdoor watermarking},
  author={Li, Yiming and Zhu, Mingyan and Yang, Xue and Jiang, Yong and Wei, Tao and Xia, Shu-Tao},
  journal={IEEE Transactions on Information Forensics and Security},
  volume={18},
  pages={2318--2332},
  year={2023},
  publisher={IEEE}
}

@article{tang2023did,
  title={Did you train on my dataset? towards public dataset protection with cleanlabel backdoor watermarking},
  author={Tang, Ruixiang and Feng, Qizhang and Liu, Ninghao and Yang, Fan and Hu, Xia},
  journal={ACM SIGKDD Explorations Newsletter},
  volume={25},
  number={1},
  pages={43--53},
  year={2023},
  publisher={ACM New York, NY, USA}
}

@article{qiao2025certdw,
  title={CertDW: Towards Certified Dataset Ownership Verification via Conformal Prediction},
  author={Qiao, Ting and Li, Yiming and Li, Jianbin and Wang, Yingjia and Qi, Leyi and Guo, Junfeng and Feng, Ruili and Tao, Dacheng},
  journal={arXiv preprint arXiv:2506.13160},
  year={2025}
}

@article{wei2024pointncbw,
  title={Pointncbw: Towards dataset ownership verification for point clouds via negative clean-label backdoor watermark},
  author={Wei, Cheng and Wang, Yang and Gao, Kuofeng and Shao, Shuo and Li, Yiming and Wang, Zhibo and Qin, Zhan},
  journal={IEEE Transactions on Information Forensics and Security},
  year={2024},
  publisher={IEEE}
}

@inproceedings{li2025towards,
  title={Towards reliable verification of unauthorized data usage in personalized text-to-image diffusion models},
  author={Li, Boheng and Wei, Yanhao and Fu, Yankai and Wang, Zhenting and Li, Yiming and Zhang, Jie and Wang, Run and Zhang, Tianwei},
  booktitle={IEEE S\&P},
  year={2025}
}

@article{li2025cbw,
  title={CBW: Towards Dataset Ownership Verification for Speaker Verification via Clustering-based Backdoor Watermarking},
  author={Li, Yiming and Yan, Kaiying and Shao, Shuo and Zhai, Tongqing and Xia, Shu-Tao and Qin, Zhan and Tao, Dacheng},
  journal={arXiv preprint arXiv:2503.05794},
  year={2025}
}

@inproceedings{guo2025audio,
  title={AUDIO WATERMARK: Dynamic and Harmless Watermark for Black-box Voice Dataset Copyright Protection},
  author={Guo, Hanqing and Guo, Junfeng and Chen, Bocheng and Wang, Yuanda and Chen, Xun and Huang, Heng and Yan, Qiben and Xiao, Li},
  booktitle={USENIX Security},
  year={2025}
}

@inproceedings{Cohen2019Certified,
  title={Certified adversarial robustness via randomized smoothing},
  author={Cohen, Jeremy and Rosenfeld, Elan and Kolter, Zico},
  booktitle={ICML},
  year={2019}
}

@article{neyman1933on,
  title={IX. On the problem of the most efficient tests of statistical hypotheses},
  author={Neyman, Jerzy and Pearson, Egon Sharpe},
  journal={Philosophical Transactions of the Royal Society of London. Series A, Containing Papers of a Mathematical or Physical Character},
  volume={231},
  number={694-706},
  pages={289--337},
  year={1933},
  publisher={The Royal Society London}
}

@inproceedings{ye2020safer,
  title={SAFER: A structure-free approach for certified robustness to adversarial word substitutions},
  author={Ye, Mao and Gong, Chengyue and Liu, Qiang},
  booktitle={ACL},
  year={2020}
}

@inproceedings{zhang2024text,
  title={Text-crs: A generalized certified robustness framework against textual adversarial attacks},
  author={Zhang, Xinyu and Hong, Hanbin and Hong, Yuan and Huang, Peng and Wang, Binghui and Ba, Zhongjie and Ren, Kui},
  booktitle={IEEE S\&P},
  year={2024}
}

@inproceedings{sun2024crowd,
  title={CROWD: Certified Robustness via Weight Distribution for Smoothed Classifiers against Backdoor Attack},
  author={Sun, Siqi and Sen, Procheta and Ruan, Wenjie},
  booktitle={EMNLP},
  year={2024}
}

@inproceedings{vovk2012conditional,
  title={Conditional validity of inductive conformal predictors},
  author={Vovk, Vladimir},
  booktitle={ACML},
  year={2012}
}

@inproceedings{liu2017neural,
  title={Neural trojans},
  author={Liu, Yuntao and Xie, Yang and Srivastava, Ankur},
  booktitle={ICCD},
  year={2017}
}

@inproceedings{liu2018fine,
  title={Fine-pruning: Defending against backdooring attacks on deep neural networks},
  author={Liu, Kang and Dolan-Gavitt, Brendan and Garg, Siddharth},
  booktitle={RAID},
  year={2018}
}

@article{el2021automatic,
  title={Automatic text summarization: A comprehensive survey},
  author={El-Kassas, Wafaa S and Salama, Cherif R and Rafea, Ahmed A and Mohamed, Hoda K},
  journal={Expert systems with applications},
  volume={165},
  pages={113679},
  year={2021},
  publisher={Elsevier}
}

@inproceedings{nasr2019comprehensive,
  title={Comprehensive privacy analysis of deep learning: Passive and active white-box inference attacks against centralized and federated learning},
  author={Nasr, Milad and Shokri, Reza and Houmansadr, Amir},
  booktitle={2019 IEEE symposium on security and privacy (SP)},
  pages={739--753},
  year={2019},
  organization={IEEE}
}

@article{li2025rethinking,
  title={Rethinking data protection in the (generative) artificial intelligence era},
  author={Li, Yiming and Shao, Shuo and He, Yu and Guo, Junfeng and Zhang, Tianwei and Qin, Zhan and Chen, Pin-Yu and Backes, Michael and Torr, Philip and Tao, Dacheng and others},
  journal={arXiv preprint arXiv:2507.03034},
  year={2025}
}

@article{li2025move,
  title={Move: Effective and harmless ownership verification via embedded external features},
  author={Li, Yiming and Zhu, Linghui and Jia, Xiaojun and Bai, Yang and Jiang, Yong and Xia, Shu-Tao and Cao, Xiaochun and Ren, Kui},
  journal={IEEE Transactions on Pattern Analysis and Machine Intelligence},
  year={2025}
}

@article{liu2023watermarking,
  title={Watermarking text data on large language models for dataset copyright},
  author={Liu, Yixin and Hu, Hongsheng and Chen, Xun and Zhang, Xuyun and Sun, Lichao},
  journal={arXiv preprint arXiv:2305.13257},
  year={2023}
}

@article{guo2024zeromark,
  title={Zeromark: Towards dataset ownership verification without disclosing watermark},
  author={Guo, Junfeng and Li, Yiming and Chen, Ruibo and Wu, Yihan and Huang, Heng and others},
  journal={NeurIPS},
  year={2024}
}

@inproceedings{bansal2022certified,
  title={Certified neural network watermarks with randomized smoothing},
  author={Bansal, Arpit and Chiang, Ping-yeh and Curry, Michael J and Jain, Rajiv and Wigington, Curtis and Manjunatha, Varun and Dickerson, John P and Goldstein, Tom},
  booktitle={ICML},
  year={2022}
}

@inproceedings{jiang2023ipcert,
  title={Ipcert: Provably robust intellectual property protection for machine learning},
  author={Jiang, Zhengyuan and Fang, Minghong and Gong, Neil Zhenqiang},
  booktitle={ICCV},
  year={2023}
}

@inproceedings{ren2023dimension,
  title={Dimension-independent certified neural network watermarks via mollifier smoothing},
  author={Ren, Jiaxiang and Zhou, Yang and Jin, Jiayin and Lyu, Lingjuan and Yan, Da},
  booktitle={ICML},
  year={2023}
}

@inproceedings{cui2022unified,
  title={A unified evaluation of textual backdoor learning: Frameworks and benchmarks},
  author={Cui, Ganqu and Yuan, Lifan and He, Bingxiang and Chen, Yangyi and Liu, Zhiyuan and Sun, Maosong},
  booktitle={NeurIPS},
  year={2022}
}

@inproceedings{pei2023textguard,
  title={Textguard: Provable defense against backdoor attacks on text classification},
  author={Pei, Hengzhi and Jia, Jinyuan and Guo, Wenbo and Li, Bo and Song, Dawn},
  booktitle={NDSS},
  year={2024}
}

\appendix

\section{Proof of Theorem \ref{thm:robustness_condition}}
\label{sec:appa_the}

We hereby provide the detailed proof of Theorem~\ref{thm:robustness_condition}. 
To establish the theorem, it suffices to verify the robustness guarantees in both the embedding and permutation spaces. 
The embedding-space proof follows directly from CertDW~\cite{qiao2025certdw}, 
so we focus on deriving the corresponding result in the permutation space. 
The proof proceeds by establishing a sequence of lemmas that progressively lead to the desired robustness condition.


\begin{lemma}[Lipschitz Continuity of Smoothed Classifier \cite{zhang2024text}]
 \label{lem:lipschitz}
Given a uniform distribution noise $\bm{\rho}\sim\mathcal{U}[-\lambda,\lambda]$, the smoothed classifier $\tilde{h}(\bm{U},\bm{W})=\mathbb{P}_{(\bm{\rho}, \bm{\varepsilon})}\!\big[h(\psi(\bm{U},\bm{\rho})\cdot W)\big]$ is $1/2\lambda$-Lipschitz in $\bm{U}$ under the $\ell_{1}$ norm.
\end{lemma}

With lemma \ref{lem:lipschitz} established, we now derive the robustness condition for dataset ownership verification.

\begin{lemma}
  \label{lem:robustness_condition_per_1}
Assume that the permutation noise $\rho \sim \mathcal{U}[-\lambda,\lambda]$. 
Denote by $W\!\big(f_{\bm\theta},\mathcal{P}_{\rho}\big)$ the watermark robustness (WR) of the suspicious model $f_{\bm\theta}$ under permutation-space smoothing, as defined in Eq.~(\ref{eq:WR}), which is $L$-Lipschitz continuous under the $\|\cdot\|_1$ norm. Let $\delta_p$ represent the permutation-space watermark perturbations, and define
$r_p=\max_{k=1,\ldots,K}\|\delta_p\|_1$ as the maximum perturbation magnitude in this space. Then, text dataset ownership verification in the permutation space is guaranteed if 
\begin{equation}
\label{eq:robudtness_condition_per}
W\!\big(f_{\bm\theta},\mathcal{P}_{\rho}\big) 
> L.r_p 
+ P_C^{(J-m-\lfloor \alpha_0(J-m+1)\rfloor)}(g_w, \mathcal{P}_{\rho}),
\end{equation}
where $P_C^{(j)}(g_w, \mathcal{P}_{\rho})$ denotes the $j$-th smallest element of the calibration set 
$P_C(\mathcal{P}_{\rho})$ constructed from benign models $g_w$. 
Here, $\alpha_0$, $J$, and $m$ are defined as in Proposition \ref{pro:conformal_predition}.  
\end{lemma}

\begin{proof}
We first define two WR quantities under permutation-space smoothing. The watermark robustness evaluated on the watermarked samples is $W\!\big(f_{\bm\theta},\mathcal{P}_{\rho}\big)=
\min_{k=1,\ldots,K}\mathbb{P}_{\rho}\!\Big(\arg\max f\big(\psi(\widehat{\bm U}_k,\rho)\cdot \bm W_k;\bm\theta\big)=\hat y\Big)$, where $\widehat{\bm U}_k$ denotes the watermarked samples. We further define the corresponding WR evaluated on the \emph{perturbed} samples as $\hat{W}\!\big(f_{\bm\theta},\mathcal{P}_{\rho}\big)=\min_{k=1,\ldots,K}\mathbb{P}_{\rho}\!\Big(\arg\max f\big(\psi({\bm U}_k,\rho)\cdot \bm W_k;\bm\theta\big)=\hat y \Big)$, where the perturbation is bounded by $\|\widehat{\bm U}_k-{\bm U}_k\|_1=\|\delta_p\|_1\le r_p$.

According to Proposition~\ref{pro:conformal_predition},  dataset ownership verification holds if the following condition is satisfied:
\begin{equation}
\label{eq:dov_condition}
\hat{W}\!\big(f_{\bm\theta},\mathcal{P}_{\rho}\big)>P_C^{(J-m-\lfloor \alpha_0(J-m+1)\rfloor)}(g_w,\mathcal{P}_{\rho}).
\end{equation}

Since the WR functional is $L$-Lipschitz continuous w.r.t.\ the input under the $\|\cdot\|_1$ norm and
$\|\widehat{\bm U}_k-{\bm U}_k\|_1\le r_p$, we have
\begin{equation}
\label{eq:lipschitz_bridge}
\hat{W}\!\big(f_{\bm\theta},\mathcal{P}_{\rho}\big)
\ge W\!\big(f_{\bm\theta},\mathcal{P}_{\rho}\big)-L\,r_p .
\end{equation}

Therefore, a sufficient condition for Eq.~(\ref{eq:dov_condition}) is
\[W\!\big(f_{\bm\theta},\mathcal{P}_{\rho}\big)-L\,r_p>P_C^{(J-m-\lfloor \alpha_0(J-m+1)\rfloor)}(g_w,\mathcal{P}_{\rho}),
\]
which is equivalent to Eq.~(\ref{eq:robudtness_condition_per}).
This completes the proof.
\end{proof}

\begin{lemma}[\textbf{Certified Watermark Robustness in Permutation Space}]
 \label{lem:robustness_condition_per}
Assume that the permutation noise $\rho \sim \mathcal{U}[-\lambda,\lambda]$. 
Denote by $W\!\big(f_{\bm\theta},\mathcal{P}_{\rho}\big)$ the watermark robustness (WR) of the suspicious model $f_{\bm\theta}$ under permutation-space smoothing, as defined in Eq.~(\ref{eq:WR}). Let $\delta_p$ represent the permutation-space watermark perturbations, and define
$r_p=\max_{k=1,\ldots,K}\|\delta_p\|_1$ as the maximum perturbation magnitude in this space. Then, text dataset ownership verification in the permutation space is guaranteed if 
\begin{equation}
\label{eq:robudtness_per}
W\!\big(f_{\bm\theta},\mathcal{P}_{\rho}\big) 
> \frac{r_p}{2\lambda} 
+ P_C^{(J-m-\lfloor \alpha_0(J-m+1)\rfloor)}(g_w, \mathcal{P}_{\rho}),
\end{equation}
where $P_C^{(j)}(g_w, \mathcal{P}_{\rho})$ denotes the $j$-th smallest element of the calibration set 
$P_C(\mathcal{P}_{\rho})$ constructed from benign models $g_w$. 
Here, $\alpha_0$, $J$, and $m$ are defined as in Proposition \ref{pro:conformal_predition}.
 
\end{lemma}

Finally, Lemma \ref{lem:robustness_condition_per} follows directly from Lemma \ref{lem:lipschitz} and Lemma \ref{lem:robustness_condition_per_1}; combining it with the embedding-space result from CertDW~\cite{qiao2025certdw} completes the proof of Theorem \ref{thm:robustness_condition}.


\vspace{0.5em}
\noindent\textbf{Proof of Theorem~\ref{thm:robustness_condition}.}
According to Lemma~\ref{lem:lipschitz}, the uniform-based smoothed classifier 
$\tilde{h}(\bm{U},\bm{W})$ is $\tfrac{1}{2\lambda}$-Lipschitz in $\bm{U}$ under the $\ell_1$ norm. 
Combining this Lemma \ref{lem:lipschitz} with Lemma~\ref{lem:robustness_condition_per_1}, and substituting 
$L=\tfrac{1}{2\lambda}$ into Eq.~(\ref{eq:robudtness_condition_per}), we obtain 
Eq.~(\ref{eq:robudtness_per}) in Lemma \ref{lem:robustness_condition_per}. Together with the embedding-space result established in CertDW~\cite{qiao2025certdw}, 
this completes the proof of Theorem~\ref{thm:robustness_condition}. 
\qed

\section{Related Works}
\subsection{Backdoor Attacks in PLMs}
\label{plms_backdoor}
Text classification is a fundamental task in natural language processing. With the prevalence of pre-trained language models (PLMs), such as BERT and GPT series, fine-tuning on downstream datasets has become the dominant paradigm for adapting models to specific tasks. However, this paradigm also amplifies the security risks of deep learning models, with backdoor attacks emerging as a notable threat \cite{gu2019badnets,li2022backdoor}. Backdoor attacks inject triggers during training or fine-tuning, enabling the model to perform well on normal inputs but produce attacker-specified outputs once the triggers appear. In text scenarios, attackers must embed triggers while preserving semantic fluency, which is more challenging than pixel-level perturbations in the vision domain. Nevertheless, the strong modeling and transferability capabilities of PLMs make them particularly vulnerable to such attacks. Consequently, backdoor attacks not only pose severe security threats but also demonstrate unique potential for dataset copyright protection \cite{li2022untargeted,guo2023domain}. Among existing approaches, data-poisoning attacks are particularly critical: by injecting triggers into training data and modifying labels, any PLM fine-tuned on the poisoned dataset will inevitably inherit the backdoor behaviors, thus enabling ownership verification \cite{liu2023watermarking}. Typical forms include word-level attacks (BadWord \cite{kurita2020weight,chen2021badnl}), which insert meaningless character triggers; sentence-level attacks (AddSent \cite{dai2019backdoor}), which append fixed sentences; and structure-level attacks (SynBkd \cite{qi2021hidden}), which exploit specific syntactic structures. In this work, we focus on BadWord and AddSent methods to construct practical watermarking techniques for text dataset copyright protection.

\subsection{Randomized Smoothing}
\label{sec:RS_app}

We here provide the formal definition of randomized smoothing (RS) \cite{Cohen2019Certified}, a certified defense technique that leverages Gaussian perturbations to provide provable robustness guarantees.
Specifically, for any input $\bm{x}$, a smoothed classifier $g$ is constructed by adding Gaussian noise $\bm{\epsilon} \sim \mathcal{N}(0,\sigma^2 I)$, where $\sigma$ is a hyperparameter controlling the robustness level. This classifier outputs the class most frequently predicted by the base model $f$ under noise perturbations: $g(\bm{x}) \triangleq \arg\max_{y \in \mathcal{Y}} \mathbb{P} (f(\bm{x}+\bm{\epsilon})=y)$. Based on the Neyman-Pearson Lemma \cite{neyman1933on}, Cohen \etal \cite{Cohen2019Certified} established the certification mechanism of this method under $\ell_2$ perturbations. Let $y_A$ denote the most probable class under noise conditions (\ie, $y_A=\arg\max_y \mathbb{P} (f(\bm{x}+\bm{\epsilon})=y)$), with corresponding probability denoted as $P_A=\mathbb{P} (f(\bm{x} + \bm{\epsilon}) = y_A)$; the maximum probability of the runner-up class is denoted as $P_B=\max_{y \neq y_A} \mathbb{P} (f(\bm{x} + \bm{\epsilon}) = y)$. Theoretical analysis shows that when an adversarial perturbation $\bm{\tau}$ satisfies $\|\bm{\tau}\|_2<r$, the classification result remains stable (\ie, $g(\bm{x}+\bm{\tau})=y_A$), where the certified radius $r$ is computed as:
\begin{equation}
\label{eq:RS_radius}
r \triangleq \frac{\sigma}{2} \left(\Phi^{-1}(P_A)-\Phi^{-1}(P_B)\right),
\end{equation}
where $\Phi^{-1}$ is the inverse of the standard Gaussian CDF.

Recently, randomized smoothing has been extended to NLP tasks for certified robustness. SAFER \cite{ye2020safer} and Text-CRS \cite{zhang2024text} provide certified defenses against adversarial attacks during inference. 
CROWD \cite{sun2024crowd} applies randomized smoothing to achieve certified robustness against backdoor attacks, aiming to neutralize malicious triggers during training. However, this goal fundamentally conflicts with dataset watermarking, which requires watermark triggers to remain effective throughout training for reliable ownership verification, and we provide a detailed comparison with certified adversarial and backdoor robustness methods in  Appendix ~\ref{Com_certified}.

\subsection{Overview of CertDW}
\label{sec:cerdw}
For a $K$-class classification task, the base classifier $f_\theta$ is trained on the protected watermarked dataset $\mathcal{D}_w$ with class-dependent perturbations $\bm{\Delta}_k$. Under the RS method, the smoothed verifier is defined as $g(\hat{\bm{x}}_k, \sigma) \triangleq \arg\max_{y} \mathbb{P}_{\bm{\epsilon}} ( f_\theta(\hat{\bm{x}}_k + \bm{\epsilon}) = y)$, where $\hat{\bm{x}}_k = \bm{x}_k + \bm{\Delta}_k$ and $\bm{\epsilon} \sim \mathcal{N}(0,\sigma^2 I)$. To capture the prediction stability of benign models under noise, the \textit{principal probability (PP)} is defined as $P(g_w, \bm{\epsilon}) = \big\| \tfrac{1}{K} \sum_{k=1}^K p(\bm{x}_k \mid g_w, \bm{\epsilon}) \big\|_\infty$, where $p_y(\bm{x} \mid g_w, \bm{\epsilon}) = \mathbb{P}_{\bm{\epsilon}} (\arg\max g(\bm{x}+\bm{\epsilon};w) = y)$, which in practice is estimated via Monte Carlo sampling. To measure watermark persistence, the \textit{watermark robustness (WR)} is defined as $W(f_\theta, \bm{\epsilon}) = \min_{k=1,\cdots,K} \mathbb{P}_{\bm{\epsilon}} ( f_\theta(\bm{x}_k + \bm{\Delta}_k + \bm{\epsilon}) = \hat{y} )$, where $\hat{y}$ is the target watermark label. The maximum watermark perturbation magnitude during training is given by $r = \max_{k=1,\dots,K} \|\bm{\Delta}_k\|_2$. From $J$ benign models, a set of PP values $\{P_C^1, \dots, P_C^J\}$ is obtained and sorted as $\{P_C^{(1)} \le \cdots \le P_C^{(J)}\}$; with allowance for $m$ outliers and significance level $\alpha_0$, the calibration threshold is then determined by conformal prediction \cite{vovk2012conditional} as $P_C^{(J-m-\lfloor \alpha_0(J-m+1)\rfloor)}(g_w, \bm{\epsilon})$. Finally, dataset ownership verification is guaranteed if and only if: $W(f_\theta, \bm{\epsilon}) > \Phi\left(\frac{r}{\sigma}\right) + P_C^{(J-m-\lfloor\alpha_0(J-m+1)\rfloor)}(g_w, \bm{\epsilon})$, where $\Phi(\cdot)$ is the CDF of the standard Gaussian distribution.

\section{Comparison with Certified Adversarial and Backdoor Robustness}
\label{Com_certified}
\subsection{Certified Backdoor Robustness}
Although both certified backdoor robustness and certified dataset watermarking involve backdoor-like triggers, they pursue fundamentally different objectives. Certified backdoor robustness focuses on ensuring that a model’s predictions remain stable even when poisoned samples appear during training or testing. In this setting, the backdoor is regarded as a harmful perturbation, and the goal is to eliminate or suppress its influence through robust or certified training \cite{pei2023textguard,sun2024crowd}. In contrast, certified dataset watermarking deliberately embeds a trigger for legitimate ownership verification. The objective is not to defend against backdoors but to preserve the effect of the trigger under bounded perturbations so that the watermark remains verifiable. Furthermore, existing certified backdoor robustness methods typically depend on modifying the training process, such as applying adversarial or randomized smoothing training to achieve formal guarantees. These approaches assume full control of model training, which is unrealistic in dataset ownership verification scenarios where the adversary or an untrusted party performs the training. Hence, certified backdoor robustness cannot be directly applied to our setting.

\begin{figure}[ht!]
    \vspace{-1em}
    \centering
    \includegraphics[width=0.48\textwidth]{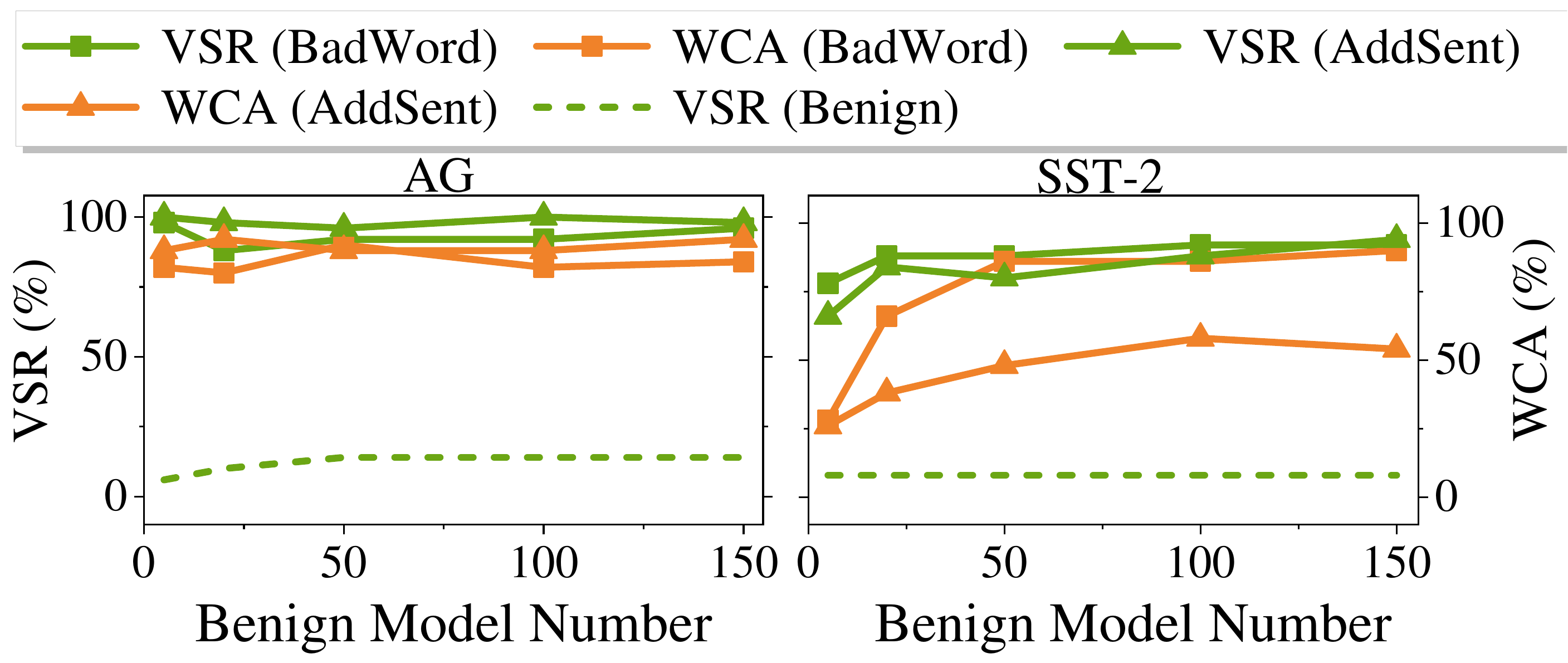}
    \vspace{-2em}
    \caption{Effects of the number of benign models.}
     \Description{Line plot showing how verification performance varies with the number of benign reference models.}
    \vspace{-2em}
    \label{fig:benign}   
\end{figure}

\subsection{Certified Adversarial Robustness}
Certified adversarial robustness and certified dataset watermarking are related in that both aim to provide provable guarantees under bounded perturbations. However, their purposes and perspectives are fundamentally different.
Certified adversarial robustness focuses on ensuring that a model’s prediction remains unchanged within a predefined neighborhood of each input, thereby characterizing the local stability of the decision boundary \cite{ye2020safer,zhang2024text}. In contrast, certified dataset watermarking investigates how a model behaves on watermarked samples that have been intentionally embedded into the training data.
Its objective is to determine whether a model has been trained on a protected dataset by measuring the stability of watermark activation rather than classification consistency. Thus, while adversarial robustness seeks insensitivity to perturbations, certified watermarking relies on persistence of specific trigger effects. Moreover, existing certified adversarial robustness methods rely on perturbation-aware model training or certified smoothing during model optimization.
These methods assume that the defender can modify or retrain the model, which is infeasible in dataset ownership verification scenarios where the suspicious model is already trained by an untrusted party. Therefore, certified adversarial robustness also cannot be directly applied to the DOV problem studied in this paper.

\section{Detailed Experimental Settings}
\label{sec:settings_app}
\subsection{Dataset Watermarking}
We follow prior work \cite{li2023black} and adopt two text backdoor watermarking methods: BadWord \cite{kurita2020weight,chen2021badnl} and AddSent \cite{dai2019backdoor}. The target label is fixed as $\hat{y}=1$, with a watermark rate of 10\%. For BadWord, we randomly select 2 triggers from {“cf”, “mn”, “bb”, “tq”, “mb”} and insert them following distributed or clustered strategies; for AddSent, we use the phrase “I watch this 3D movie” at sentence start, middle, end, or a random position. To constrain stealthiness, $\ell_2$ perturbations are kept around 0.6 (BadWord) and 1.0 (AddSent) by scaling trigger embeddings with local average pooling over adaptive windows ($w=2$ if $n<10$, $w=10$ if $n\geq80$). Token-level permutations are added during training for robustness. For each method–dataset pair, we train 50 watermarked models.

\subsection{Dataset Verification}
We reserve 5,000 test samples from AG and SST-2 as verification sets. We train $J=100$ benign models and construct calibration sets with outlier filtering ratios $\beta=0.05$ (AG) and $\beta=0.2$ (SST-2). To estimate false positives, we train 50 independent models on clean data. Verification uses dual-space perturbations: Gaussian noise in the embedding space and local window shuffling in the permutation space. For each test sample, we generate $M=1024$ perturbed versions, computing watermark robustness (WR) and principal probability (PP) via Monte Carlo estimation. The significance level $\alpha_0$ is set to 0.05, ensuring 95\% confidence under conformal prediction.

\section{The Analysis of Computational Complexity}

In this section, we analyze the computational complexity of the proposed DDSmoothing method, focusing on the stages of watermarked dataset generation and ownership verification.

\vspace{0.3em}
\noindent \textbf{Complexity of Watermarked Dataset Generation.}
Let $N$ denote the size of the text dataset and $\gamma$ the poisoning rate. In the generation stage, we only need to embed watermarks into $\gamma N$ samples, resulting in a complexity of $\mathcal{O}(\gamma N)$. Moreover, we introduce perturbations in both the embedding and permutation spaces: in the embedding space, we optimize trigger embeddings via local pooling and iterative scaling, with a complexity of $\mathcal{O}(|P|\cdot u \cdot d)$, where $|P|$ is the number of trigger positions, $u$ is the window size, and $d$ is the embedding dimension; in the permutation space, we perform local group-based reordering for a sequence of length $n$, which incurs a complexity of $\mathcal{O}(n)$. Since these two components impose relatively minor overhead compared to the overall data processing scale, the total complexity can still be regarded as $\mathcal{O}(\gamma N)$, with negligible additional runtime overhead.

\vspace{0.3em}
\noindent \textbf{Complexity of Ownership Verification.}
Let $J$, $W_a$, and $I_n$ denote the numbers of benign models, watermarked models, and independent models, respectively. Training these models incurs complexities of $\mathcal{O}(J)$, $\mathcal{O}(W_a)$, and $\mathcal{O}(I_n)$, and can be significantly accelerated via parallel training. During verification, we compute the watermark robustness (WR) and the principal probability (PP) under dual-space randomized smoothing. For the suspicious model, WR is estimated by running $M$ Monte Carlo trials on a watermarked example from each class while injecting noise into both the embedding and permutation spaces, resulting in a complexity of $\mathcal{O}(K \cdot M \cdot \mathcal{C}(f))$, where $K$ is the number of classes and $\mathcal{C}(f)$ denotes the cost of a single forward pass. For benign models, PP is computed by estimating prediction distributions across $J$ models, with a complexity of $\mathcal{O}(J \cdot K \cdot M \cdot \mathcal{C}(g))$. Since WR/PP only involve forward inference and are highly parallelizable on GPUs, their cost is typically negligible compared to model training. For example, on the AG dataset with BERT, training a benign model takes about 5 minutes, while computing WR and PP takes about 30 seconds, making the verification overhead relatively small and acceptable.

\section{Additional Results on the AG}
\label{appendix:noise_adv}
\begin{figure}[!t]
    \centering
     \vspace{-0.5em}
    \subfigure[AG (BadWord)]{
		\includegraphics[width=0.228\textwidth]{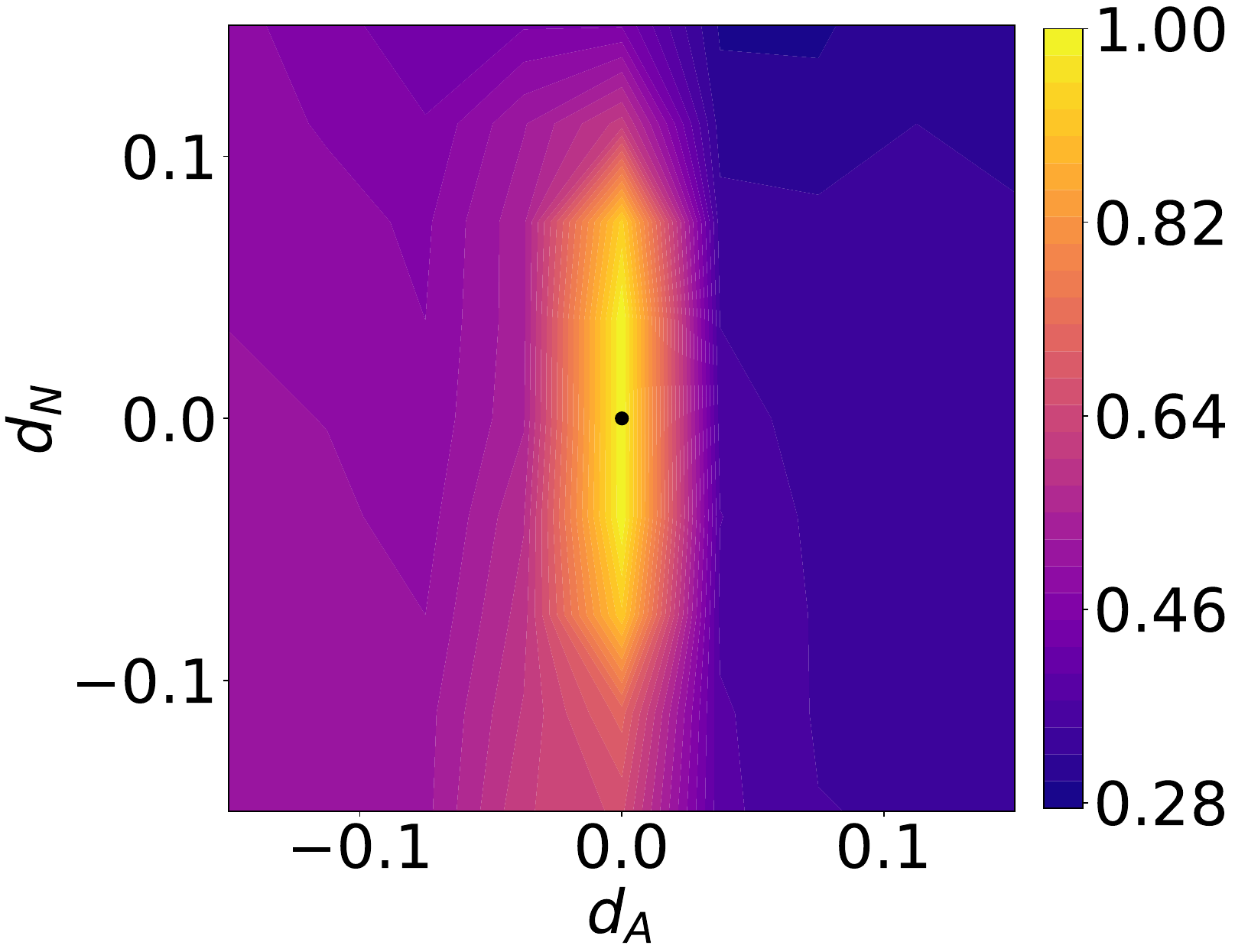}}
        \hspace{0.1em}
    \subfigure[AG (AddSent)]{
		\includegraphics[width=0.228\textwidth]{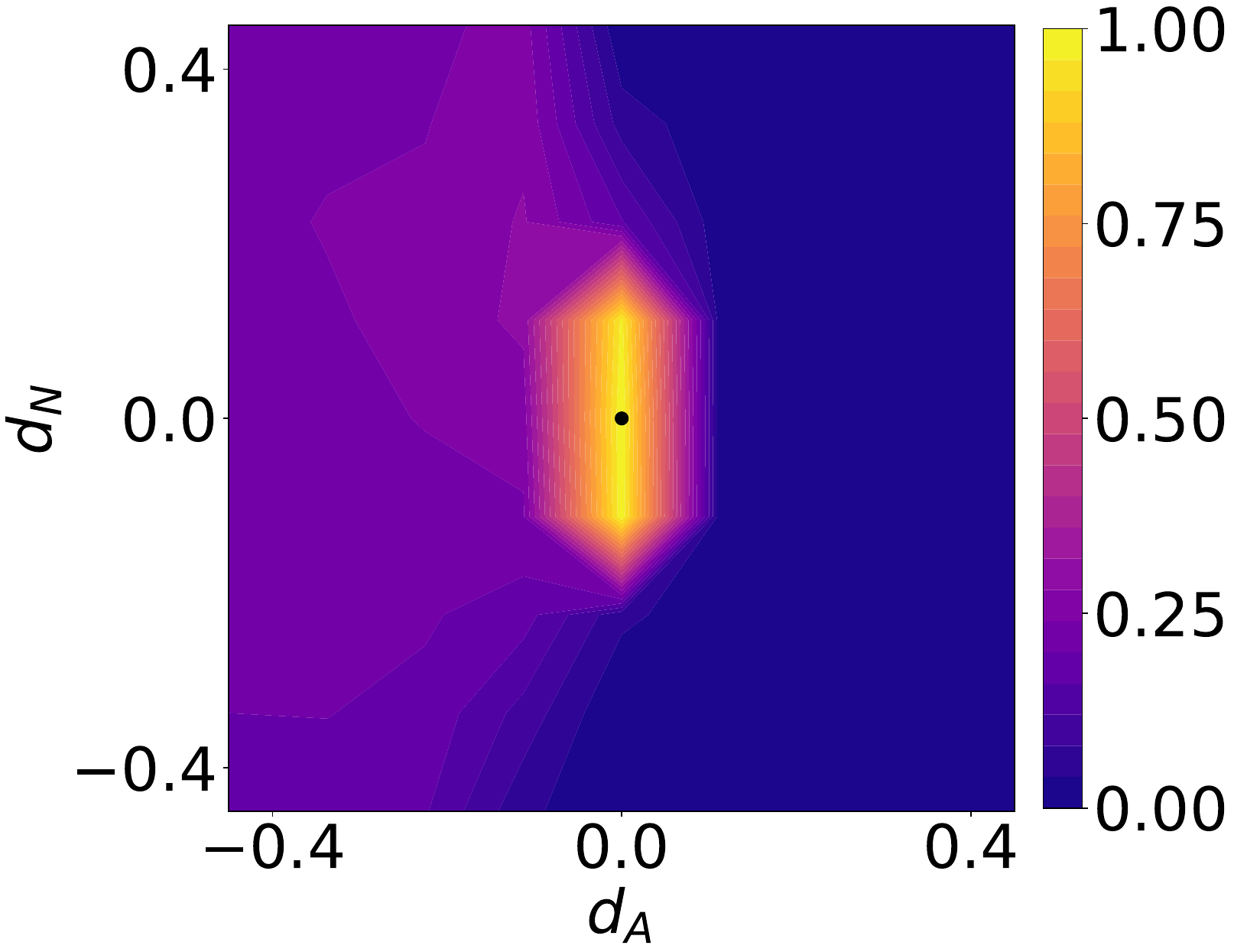}} 
        \vspace{-1.8em}
   \caption{Additional results on the AG dataset. WSR of watermarked samples under perturbations in the embedding space. $d_N$ and $d_A$ denote the noise and adversarial directions, respectively, and ‘$\bullet$’ marks the original watermarked sample. Left: BadWord; Right: AddSent.}
   \Description{Two subfigures for the AG dataset comparing watermark robustness (WR) under embedding-space perturbations. The left panel uses the BadWord trigger and the right panel uses AddSent; the bullet marks the original watermarked sample and the curves show WR under noise and adversarial directions.}
    \label{fig:noise_adv_AG}
    \vspace{-0.8em}
\end{figure}

\looseness=-1
For completeness, we present the results on the AG News dataset in Figure~\ref{fig:noise_adv_AG}. The trends are consistent with those observed on SST-2 in the main paper: while stochastic noise moderately reduces watermark performance, adversarial perturbations are substantially more destructive, further confirming the fundamental vulnerability of existing PLMs dataset watermarking methods under worst-case perturbation scenarios.

\section{Additional Effectiveness Results on the AG}
\label{appendix:effectiveness}
\begin{figure}[!t]
    \centering
     \vspace{-0.5em}
    \subfigure[AG (BadWord)]{
		\includegraphics[width=0.228\textwidth]{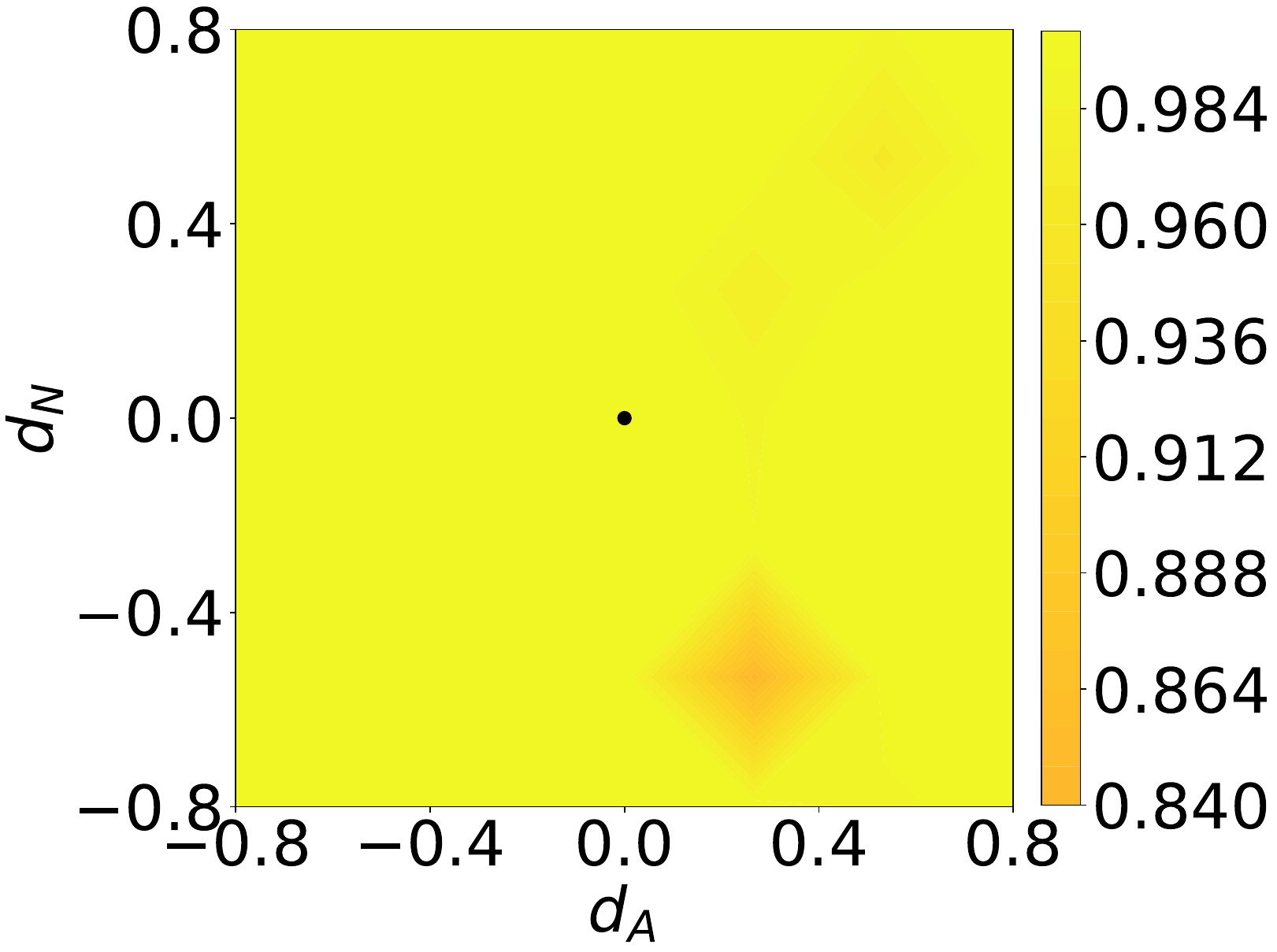}}
	\hspace{0.1em}
    \subfigure[AG (AddSent)]{
		\includegraphics[width=0.228\textwidth]{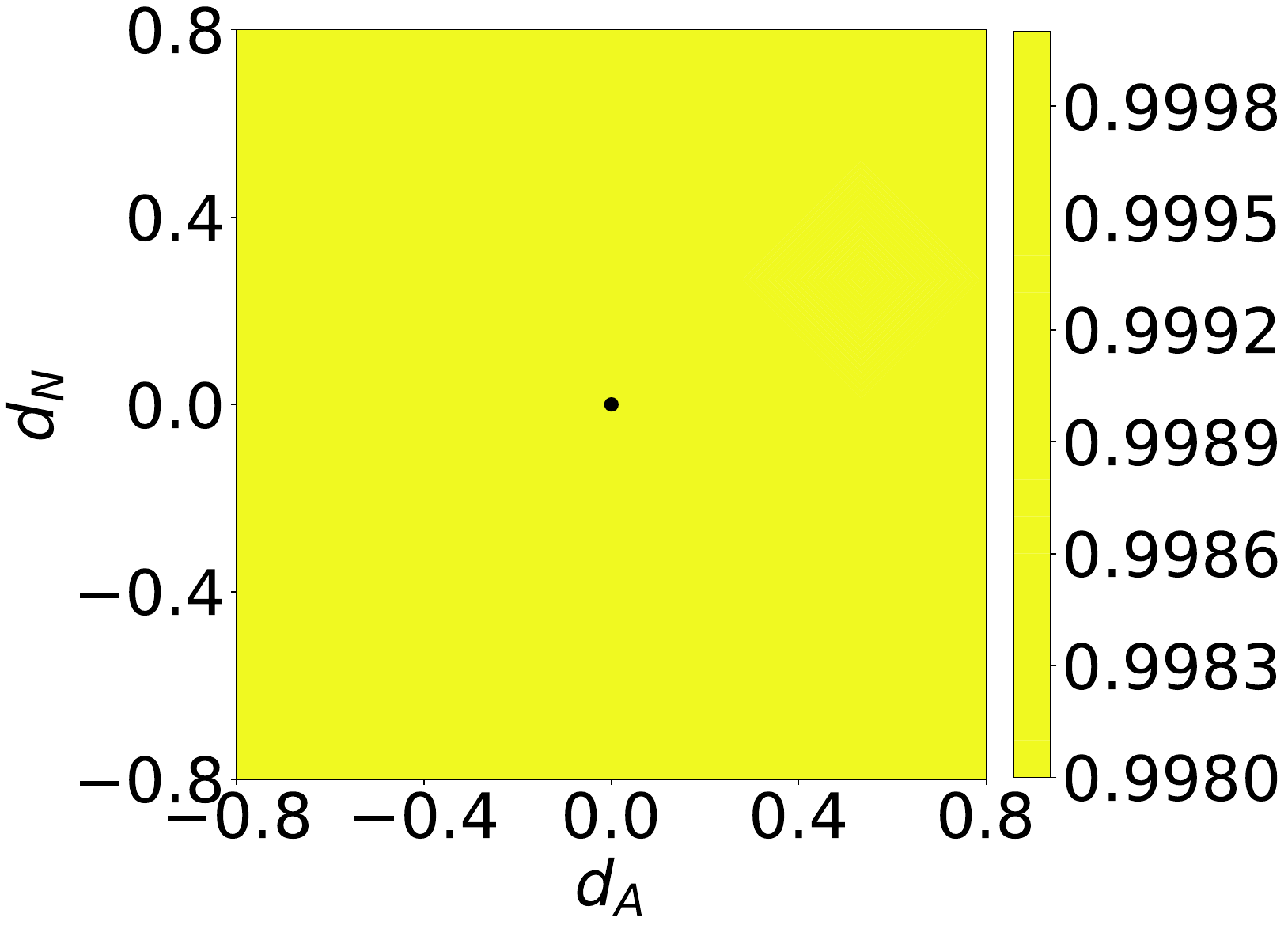}}
     \vspace{-1.5em}
    \caption{WSR of our method’s watermarked samples on the AG dataset under perturbations in the embedding space. $d_N$ and $d_A$ denote the noise and adversarial directions, respectively, and ‘$\bullet$’ marks the original watermarked sample. Left: BadWord. Right: AddSent.}
     \vspace{-0.8em}
     \Description{Two subfigures for the AG dataset comparing watermark robustness (WR) of our method under embedding-space perturbations. The left panel uses the BadWord trigger and the right panel uses AddSent; the bullet marks the original watermarked sample and the curves show WR under noise and adversarial directions.}
    \label{fig:noise_adv_ourmodel_AG}
\end{figure}

For completeness, we report additional results on the AG News dataset in Figure~\ref{fig:noise_adv_ourmodel_AG}. The results are consistent with those on SST-2, demonstrating that watermarked samples maintain nearly 99\% WSR under stochastic noise and remain robust even under worst-case perturbations, indicating that completely removing the watermark becomes significantly more difficult.

\section{Effects of the Number of Benign Models}\label{sec:abl_app}
\looseness=-1
As shown in Figure \ref{fig:benign}, both VSR and WCA increase with the number of benign models $J$. On AG, both methods maintain 80–100\%, while on SST-2, BadWord’s WCA rises from 28\% to 90\% and VSR stays above 66\%. The VSR of benign models remains around 10\%, indicating a low false positive rate. Results show that performance converges when $J$ reaches 50–100, with limited gains afterward. Thus, dataset owners should choose the number of benign models based on their needs.

\begin{figure}[!t]
    \centering
    \vspace{-1em}
    \subfigure[AG (VSR)]{
		\includegraphics[width=0.22\textwidth]{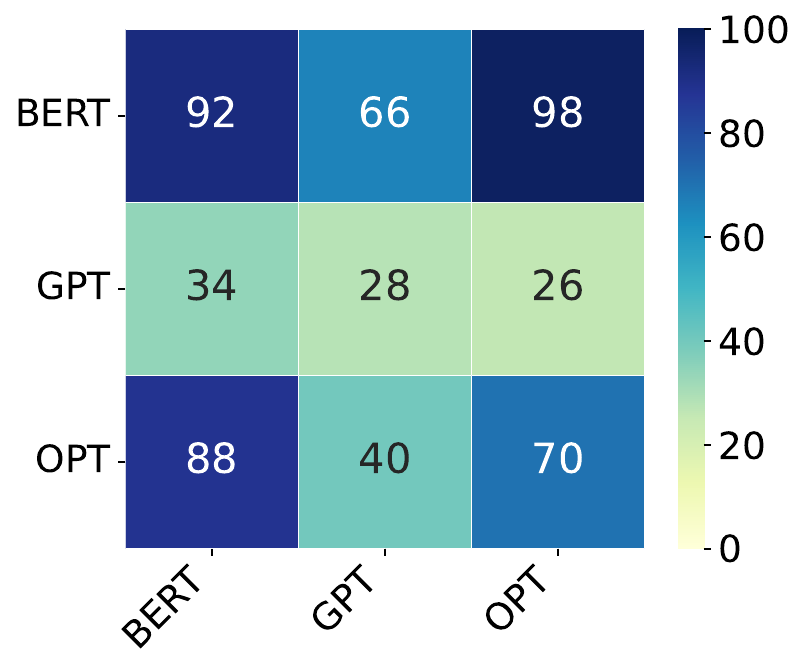}
        \label{AG_VSR}}
    \hspace{0.1em}
    \subfigure[AG (WCA)]{
		\includegraphics[width=0.22\textwidth]{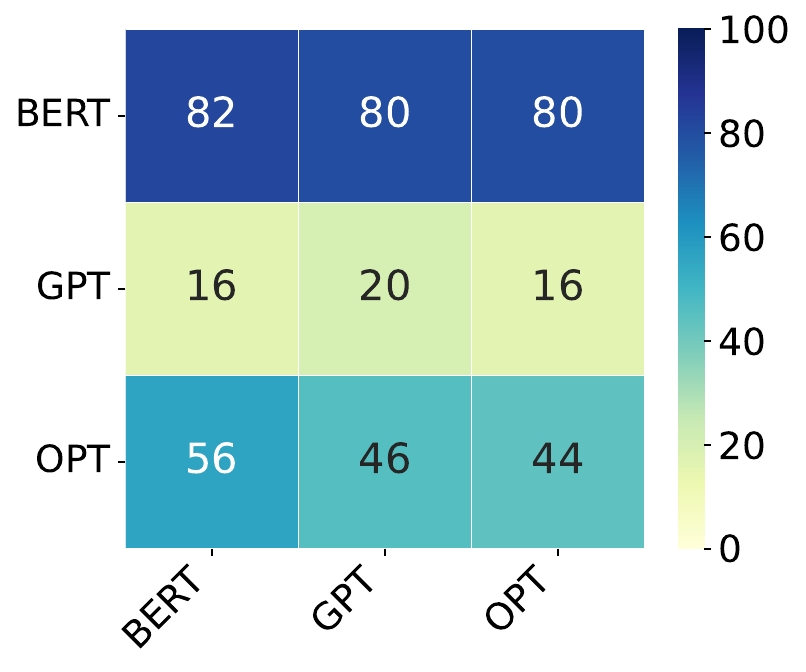}
        \label{AG_WCA}}  
	\vspace{-1.2em}
    \caption{Performance on AG with different benign (rows) and watermarked (columns) model architectures.}
    \Description{Two heatmaps for the AG dataset. The left heatmap shows verification success rate (VSR) and the right heatmap shows watermark certification accuracy (WCA), comparing different model architectures where rows correspond to benign models and columns correspond to watermarked models.}
    \label{fig:model_transfer_ag}
    	\vspace{-0.8em}
    \end{figure}

\section{Model-level Transferability on AG}
\label{sec:model_app}
Figure~\ref{fig:model_transfer_ag} shows the results on the AG dataset. The overall trends are consistent with those observed on SST-2, demonstrating that DDSmoothing remains effective across different architectures.

\end{document}